\documentclass[twocolumn,twocolappendix]{aastex631}

\usepackage[T1]{fontenc}
\usepackage{ae,aecompl}
\usepackage{mathtools,tabu}
\usepackage{amsmath,amssymb}
\usepackage{upgreek} 

\makeatletter 
\newcommand{\LCASES}[1]{$\m@th\displaystyle{#1}$\hfil}
\newcommand{\CCASES}[1]{\hfil$\m@th\displaystyle{#1}$\hfil}
\newcommand{\RCASES}[1]{\hfil$\m@th\displaystyle{#1}$}
\makeatother

\newcases{ecases}{\quad}{\CCASES{##}}{\LCASES{##}}{\lbrace}{.}
\newcases{ecases*}{\quad}{\CCASES{##}}{{##}\hfil}{\lbrace}{.}

\graphicspath{{Figures/}}

\def\beq{\begin{eqnarray}}
\def\eeq{\end{eqnarray}}

\def\Msun{M_\odot}
\def\siglnReff{\sigma_{\log r_\mathrm{eff}}}
\def\siglogM{\sigma_{\log M}}
\def\Mmin{M_\mathrm{min}}
\def\fcen{f_\mathrm{cen}}

\def\fcenO{f_\mathrm{cen,0}}
\def\Mstar{M_c}
\def\hinvMpc{h^{-1} \mathrm{Mpc}}

\newcommand{\andres}[1]{\textcolor{red}{AS: #1}}

\shorttitle{DES-Y3 Source HOD Modeling}
\shortauthors{Salcedo, Eifler \& Behroozi}

\begin{document}

\title{One Galaxy Sample to Rule Them All: Halo Occupation Distribution Modeling of DES Year 3 Source Galaxies}

\correspondingauthor{Andr\'{e}s N. Salcedo}
\email{ansalcedo@arizona.edu}

\author[0000-0003-1420-527X]{Andr\'{e}s N. Salcedo}
\affiliation{Department of Astronomy/Steward Observatory, University of Arizona, 933 North Cherry Avenue, Tucson, AZ 85721-0065, USA}
\affiliation{Department of Physics, University of Arizona, 1118 East Fourth Street, Tucson, AZ 85721, USA }
\author[0000-0002-1894-3301]{Tim Eifler}
\affiliation{Department of Astronomy/Steward Observatory, University of Arizona, 933 North Cherry Avenue, Tucson, AZ 85721-0065, USA}
\affiliation{Department of Physics, University of Arizona, 1118 East Fourth Street, Tucson, AZ 85721, USA }
\author[0000-0002-2517-6446]{Peter Behroozi}
\affiliation{Department of Astronomy/Steward Observatory, University of Arizona, 933 North Cherry Avenue, Tucson, AZ 85721-0065, USA}
\affiliation{National Astronomical Observatory of Japan, 2-21-1 Osawa, Mitaka, Tokyo, 181-8588, Japan}

\begin{abstract}
For the joint analysis of second-order weak lensing and galaxy clustering statistics, so-called $3{\times}2$ analyses, the selection and characterization of optimal galaxy samples is a major area of research. One promising choice is to use the same galaxy sample as lenses and sources, which reduces the systematics parameter space that describes uncertainties related to galaxy samples. Such a ``lens-equal-source'' analysis significantly improves self-calibration of photo-z systematics leading to improved cosmological constraints. With the aim to enable a lens-equal-source analysis on small scales we investigate the halo-galaxy connection of DES-Y3 source galaxies. We develop a technique to construct mock source galaxy populations by matching COSMOS/UltraVISTA photometry onto {\sc{UniverseMachine}} galaxies. These mocks predict a source halo occupation distribution (HOD) that exhibits significant redshift evolution, non-trivial central incompleteness and galaxy assembly bias. We produce multiple realizations of mock source galaxies drawn from the {\sc{UniverseMachine}} posterior with added uncertainties in measured DES photometry and galaxy shapes. We fit a modified HOD formalism to these realizations to produce priors on the galaxy-halo connection for cosmological analyses. We additionally train an emulator that predicts this HOD to $\sim2\%$ accuracy from redshift $z = 0.1 - 1.3$ that models the dependence of this HOD on 1) observational uncertainties in galaxy size and photometry, and 2) uncertainties in the {\sc{UniverseMachine}} predictions. 
\end{abstract}


\section{Introduction}
Observations of the large-scale structure (LSS) of the Universe contain a wealth of information on fundamental physics questions such as theories of gravity, the mass and number of species of neutrinos, and the nature of dark energy and dark matter. Existing and future galaxy redshift surveys such as the Kilo-Degree Survey \citep[KiDS,][]{2017MNRAS.465.1454H,2021A&A...646A.140H}, the Dark Energy Survey \citep[DES,][]{2018PhRvD..98d3526A,2021arXiv210513549D}, the Hyper Suprime-Cam \citep[HSC,][]{2019PASJ...71...43H}, the Baryon Oscillation Spectroscopic Survey, the Dark Energy Spectroscopic Instrument \citep[DESI,][]{2016arXiv161100036D}, the Vera C. Rubin Observatory Legacy Survey of Space and Time \citep[LSST,][]{2019ApJ...873..111I}, the Nancy Grace Roman Space Telescope \citep{2019arXiv190205569A}, the Euclid \citep{Euclid_WhitePaper}, and the Spectro-Photometer for the History of the Universe, Epoch of Reionization, and Ices Explorer \citep[SPHEREx,][]{2014arXiv1412.4872D} aim to unlock this information through a variety of imaging and spectroscopic measurements.  

One of the main challenges for joint analyses of weak lensing and galaxy clustering observables in ongoing and future surveys is the modeling of small scales. While uncertainties in the non-linear evolution of the matter density field pose a common problem for both lensing and clustering, more significant limitations arise from our lack of understanding of feedback and cooling processes that affect baryons and inadequacies in modeling the galaxy-halo connection. For cosmic shear, which is directly sensitive to the matter distribution, baryonic modeling uncertainties are the main limitation when pushing to small scales. For galaxy clustering and for galaxy-galaxy lensing, uncertainties in modeling the galaxy-halo connection are the dominant systematic that prohibit the inclusion of small scales. 

In the clustering and galaxy-galaxy lensing category a variety of concepts have been proposed to go beyond the galaxy bias description of linear perturbation theory \citep{kaiser:84}. Higher-order perturbative models for galaxy biasing are an active research area \citep[see][for a review]{2018PhR...733....1D} pushing our ability to model galaxy bias into the quasi-linear regime. More recently, this boundary has been pushed to even smaller scales by Hybrid Effective Field Theory models (HEFT), which utilize a combination of analytical bias description and numerical displacements calculated from N-body simulations to describe the statistical distribution of the galaxy density field \citep{ccw20,kdc21,hga21,zap23}. 

Halo Occupation Distribution (HOD) models fall into the category of empirical models that describe the galaxy-halo connection \citep[e.g.][]{Berlind_2002,Zehavi_et_al_2005,bmc13}. HOD methods statistically specify the relationship between galaxies and their host halos, primarily as a function of host halo mass. These methods enable interpretation of measurements of galaxy-galaxy lensing and galaxy clustering into the fully non-linear regime. This is done at the expense of introducing additional free parameters, but non-linear clustering data can constrain them leading to improved constraints on cosmological parameters \citep[e.g.][]{Yoo_et_al_2006, Zheng_Weinberg_2007, Cacciato_et_al_2009, Cacciato_et_al_2012,Cacciato_et_al_2013,Leauthaud_et_al_2011,Yoo_Seljak_2012, More_et_al_2013}.

For the design of clustering analyses of future surveys it is necessary to consider the impact of galaxy bias uncertainties on the overall error budget holistically. Statements that a specific galaxy bias parameterization is accurate at a specific $\%$-level up to a specific $k$ are interesting, but insufficient in the context of quantifying the trade space of parameter bias and errors in the posterior probability. When pushing to smaller scales, the hope of smaller confidence intervals is countered by the need for additional model complexity to reduce cosmological parameter biases. Model complexity enters through additional nuisance parameters that open up more degrees of freedom and through increasingly wide priors on the existing nuisance parameters. Both aspects affect not only the small scales that analysts aim to include, unfortunately these parameterizations impact all scales of the analysis. As a consequence the increased model complexity that allows for the inclusion of small scales can result in a degradation of constraining power overall. 

The DES Y3 joint analysis of galaxy clustering, galaxy galaxy lensing, and cosmic shear (referred to as 3x2) \citep{DESY33x2} uses linear galaxy bias with one galaxy bias parameter per tomographic bin. With respect to uncertainties in small scale bias modeling this analysis has been shown to be unbiased with scale-cuts of 6 and 8 $\hinvMpc$ for galaxy-galaxy lensing and clustering, respectively \citep{kfp21}. The DES Y3 2x2 analysis \citep{pkd22}, which only combines galaxy clustering and galaxy-galaxy lensing, explores the inclusion of smaller scales in the analysis at the cost of adding a second parameter per tomographic bin (1-loop perturbation theory) to describe non-linear galaxy bias evolution. As a result this analysis is able to include scales down to 4 and 8 $\hinvMpc$ for clustering and galaxy-galaxy lensing respectively, without incurring cosmological parameter biases. The resulting gain in cosmological information is negligible however, likely due to the increased parameter space. 

A corresponding analysis from the HSC collaboration using CMASS galaxies as the lens sample and HSC galaxies as sources is presented in \citep{smm23, mst23}. The first analysis relies on perturbation theory to push to smaller scales and the latter uses an HOD model of CMASS galaxies \citep{mmm15}. \cite{mkt22} use 4 different variations of this HOD to populate numerical simulations, generate corresponding mock catalogs, and show that their pipeline can recover the input cosmology robustly given wide priors on the HOD parameters. 

The main goal of this paper is to derive an HOD model for the DES Y3 source sample that includes realistic, informative priors on the HOD parameters. Following \cite{Schaan_et_al_2020,Fang_et_al_2022} this will enable future joint analyses of e.g. galaxy clustering, weak lensing, CMB lensing and their cross-correlations using only the source galaxy sample without the need to characterize a separate lens or clustering sample. This ``lens-equal-source'' approach is to reduce the systematics parameter space, in particular regarding photo-z parameters, and to improve self-calibration of other systematic effects. 

The source sample HOD is obtained via a matching process utilizing\textsc{UniverseMachine} \citep{Behroozi_et_al_2019} mock data and COSMOS/UltraVISTA photometry \citep{Muzzin_et_al_2013}. This process is used to generate mock catalogs of DES galaxy imaging, we then apply DES-Y3 source selection criteria and measure the resulting HOD. We subsequently vary the input parameters to our machinery (colors, magnitudes, sizes) to determine uncertainties in the source HOD, which can be used as priors in future analyses. 

The paper is structured as follows: In Section \ref{sec:mocks} we describe the construction of our DES Y3 mock catalogs, that lead to the construction of the HOD in Section \ref{sec:HOD} where we also characterize the uncertainties in our mock construction process. We derive priors on the relevant parameters in Section \ref{sec:hodpriors} and conclude in Section \ref{sec:conclusions}    .

\section{Constructing mocks of DES galaxy imaging}
\label{sec:mocks}
\subsection{Source galaxy selection in Dark Energy Survey Year 3 data}
\label{sec:source_cuts}

The parent sample for the DES-Y3 source sample \citep{Gatti_et_al_2021} is the DES-Y3 GOLD catalog (GOLD catalog from here on). The GOLD catalog comprises 326,049,983 objects over $\sim 4143 \, \mathrm{deg}^2$ of sky with observations in the {\it{griz}} bands. The {\sc{metacalibration}} algorithm \citep{Sheldon_Huff_2017, Huff_Mandelbaum_2017} was applied to this catalog using {\it{riz}}\footnote{The {\it{g}}-band was omitted due to known issues with PSF estimation} bands to produce a shape catalog from which the source sample was selected. 

To construct the source galaxy catalog the following selections were applied to the GOLD catalog,
\begin{enumerate}
\item Any objects outside of the unmasked regions of the GOLD catalog and flagged as ``anomalous'' are removed. 
\item A signal-to-noise cut of $10 < \mathrm{S/N} < 1000$ is applied. The low end of this cut removes faint objects impacted by detection biases while the high end removes very bright objects for which Poisson noise could be dominant relative to typical background noise.
\item A PSF-size-ratio cut is applied, $T / T_\mathrm{PSF} > 0.5$, where,
\beq
T = I_{xx} + I_{yy},
\eeq
referred to as the galaxy size in this context, is the sum of second moments of the Gaussian-model surface brightness profile,
\beq
I_{\mu \nu} = \frac{\int dx dy I(\mu, \nu) (\mu - \bar \mu) (\nu - \bar \nu)}{\int dx dy I(\mu, \nu)}.
\eeq
The sizes used are averaged from all exposures and bands.

\item A further size cut of $T < 10 \; \mathrm{arcsec}^2$ is applied to remove the largest objects. Visual inspection found that most of the objects cut are not themselves large but have their size estimate affected by large neighbors.
\item Objects with $T > 2 \; \mathrm{arcsec}^2$ and $\mathrm{S/N} < 30$ are also cut. These objects are relatively large and faint and are mostly blended upon visual inspection.
\item Color and magnitude cuts are applied to limit objects to those with the best photometric redshifts. These cuts are,
\begin{align}
18 < &i < 23.5, \\
15 < &r < 26, \\
15 < &z < 26, \\
-1.5 < &r - i < 4,\\
-1.5 < &z-i < 4.
\end{align}
\item A final selection was made to limit binary-star contamination. For high ellipticity objects ($|e| > 0.8$) a cut is made in {\it{r}}-T space,
\begin{equation}
    \mathrm{log}_{10}( T / \mathrm{arcsec}^2) < (22.5 - r)/2.5.
\end{equation}
\end{enumerate}
The number of objects that pass this selection is $100,204,026$. While a large fraction of GOLD catalog objects fail the various $\mathrm{S/N}$ criteria only a relatively small fraction ($\sim6\%$) require {\it{only}} $\mathrm{S/N}$ information to be excluded, i.e. are not also removed by a cut on magnitude, color, or size. 

\subsection{{\sc{UniverseMachine}} galaxies}

The {\sc{UniverseMachine}} algorithm \citep{Behroozi_et_al_2019} models the connection between galaxy and halo assembly by parametrizing galaxy star formation rates (SFRs) as a function of halo potential well depth (specifically the maximum circular velocity at the time of peak halo mass, $v_\mathrm{max}(z_{M_\mathrm{peak}}$)), redshift and assembly history with a total of 44 free model parameters. This parametrization is constrained by a variety of observational data: observed stellar mass functions, SFRs, quenched fractions, UV luminosity functions, UV-stellar mass relations, IRX-UV relations, projected auto- and cross-correlation functions, and the dependence of quenching fraction on environment. The output of this algorithm is a catalog of mock galaxies with realistic SFRs, stellar masses, and UV-luminosities as well as host subhalo properties. 

\begin{figure}
\centering \includegraphics[width=0.45\textwidth]{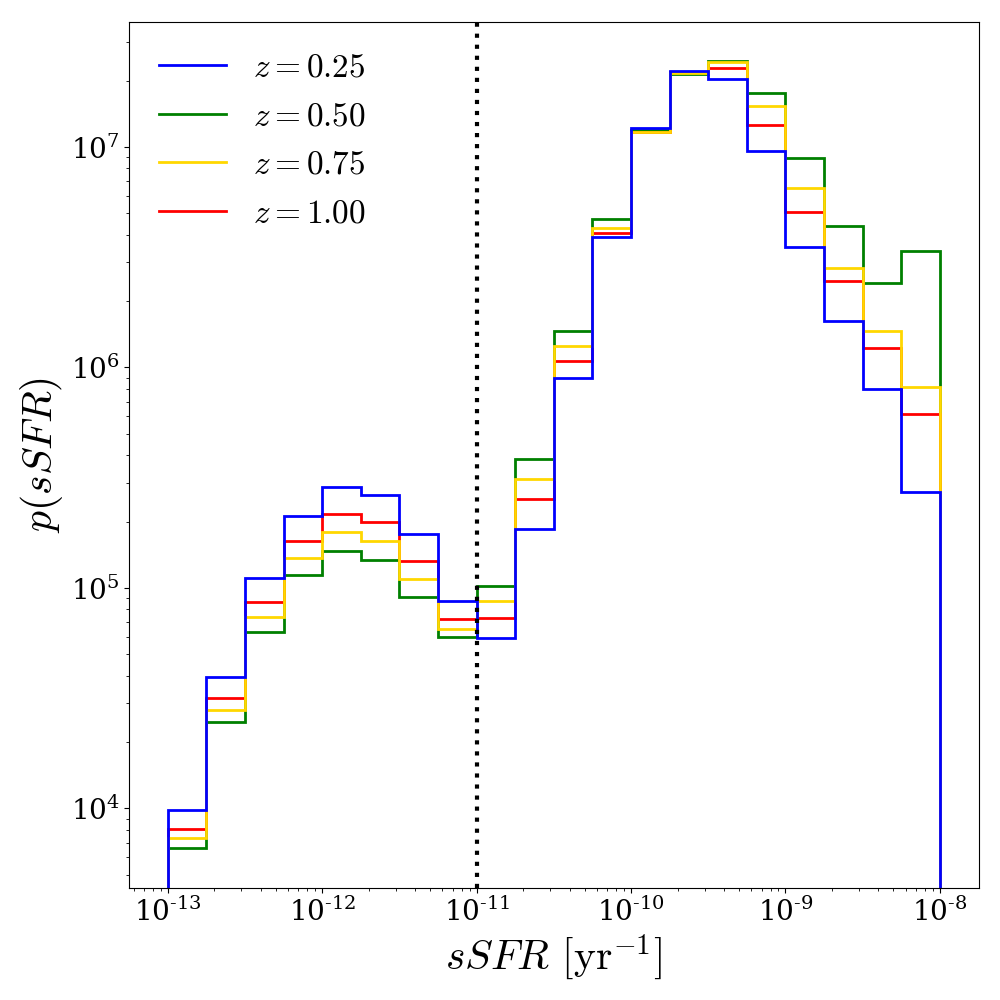}
     \caption{Distribution of specific-SFR (sSFR) in the SMDPL {\sc{UniverseMachine}} mock galaxy catalog at $z=0.25$ (blue), $z = 0.50$ (green), $z = 0.75$ (yellow), and $z=1.0$ (red). The dotted vertical line at $\mathrm{sSFR} = 10^{-11} \, \mathrm{yr}^{-1}$ located at a local minimum in the distribution divides galaxies into star-forming and quiescent. We observe that the location of this minimum is insensitive to redshift.}
\label{fig:sSFR-dist}
\end{figure}

In what follows we utilize publically available {\sc{UniverseMachine}} catalogs\footnote{https://www.peterbehroozi.com/data.html} built on the Small MultiDark Planck (SMDPL) and Bolshoi-Planck simulations \citep{Klypin_MultiDark_et_al_2016}.\footnote{https://www.cosmosim.org/cms/simulations/smdpl/} Both are dark-matter-only cosmological N-body simulations in periodic cubes, with particle resolution of $M_\mathrm{part} \sim 1{-}2 \times 10^{8} \; h^{-1} \Msun$, sufficient to resolve the host halos of DESY3 source galaxies. The SMDPL cube has side length $L_\mathrm{side} = 400 \; h^{-1} \mathrm{Mpc}$ while Bolshoi-Planck has side length $L_\mathrm{side} = 250 \; h^{-1} \mathrm{Mpc}$.The SMDPL cosmology is based on the \citet{Planck_2016} results with $\Omega_m = 0.307$, $\Omega_\Lambda = 0.692$, $\sigma_8 = 0.823$ and $h = 0.678$. The Bolshoi-Planck cosmology is the same with the exception of $h = 0.70$. In what follows we use virial halo masses $M_\mathrm{vir}$ based on the redshift-dependent spherical overdensity definition of \citet{Bryan_1998}. Fig. \ref{fig:sSFR-dist} shows the specific star-formation rate (sSFR) distribution at $z = 0.5$ and $z = 1.0$ within these catalogs. We see at both low- and high-redshift a clear splitting into a quiescent and star-forming population at $\mathrm{sSFR} = 10^{-11}\,\mathrm{yr}^{-1}$. 

\subsection{COSMOS/UltraVISTA galaxy catalog}

We use the public $K_s$-selected catalog COSMOS/UltraVISTA catalog of \citet{Muzzin_et_al_2013} to assign colors to our mock galaxies. This catalog aggregates photometric data taken in the COSMOS/UltraVISTA field. Far to near UV imaging was provided by the {\it{Galaxy Evolution Explorer}} satellite \citep[GALEX;][]{Martin_et_al_2005}, optical broad to medium band imaging from the Canada-France-Hawaii Telescope \citep[CFHT;][]{Capak_et_al_2007}, near-infrared data from the UltraVISTA survey \citep{McCracken_et_al_2012}, and mid-infrared data from {\it{Spitzer}} \citep{Sanders_et_al_2007_Spitzer}. Galaxies are identifed in this data and those with $K_s > 24.35$ are kept in the catalog. The resulting catalog covers $1.62\,\mathrm{deg}^2$ of the sky and provides photometry in 30 bands (covering a wavelength range of 0.15{-}24$\, \mu \mathrm{m}$) for 262,615 galaxies. 

\citet{Muzzin_et_al_2013} provide photometric redshifts calculated using the EAZY software \citep{Brammer_et_al_2008_EAZY}. Based on spectroscopic overlap these photometric redshifts are reliable at $z_\mathrm{spec} < 1.5$ with an error of $\delta z /( 1+z) = 0.013$, with a $1.56\%$ outlier fraction ($>3\sigma$ from the mean relation). \citet{Muzzin_et_al_2013} also provide estimates of stellar mass and SFR obtained from SED fits using either models from \citet{Bruzual_Charlot_2003} or models from \citet{Maraston_2005}. In what follows we will use stellar masses and SFR based on the \citet{Maraston_2005} models, though we have checked that this choice has no qualitative impact on our results.

\begin{figure*}
\centering \includegraphics[width=0.8\textwidth]{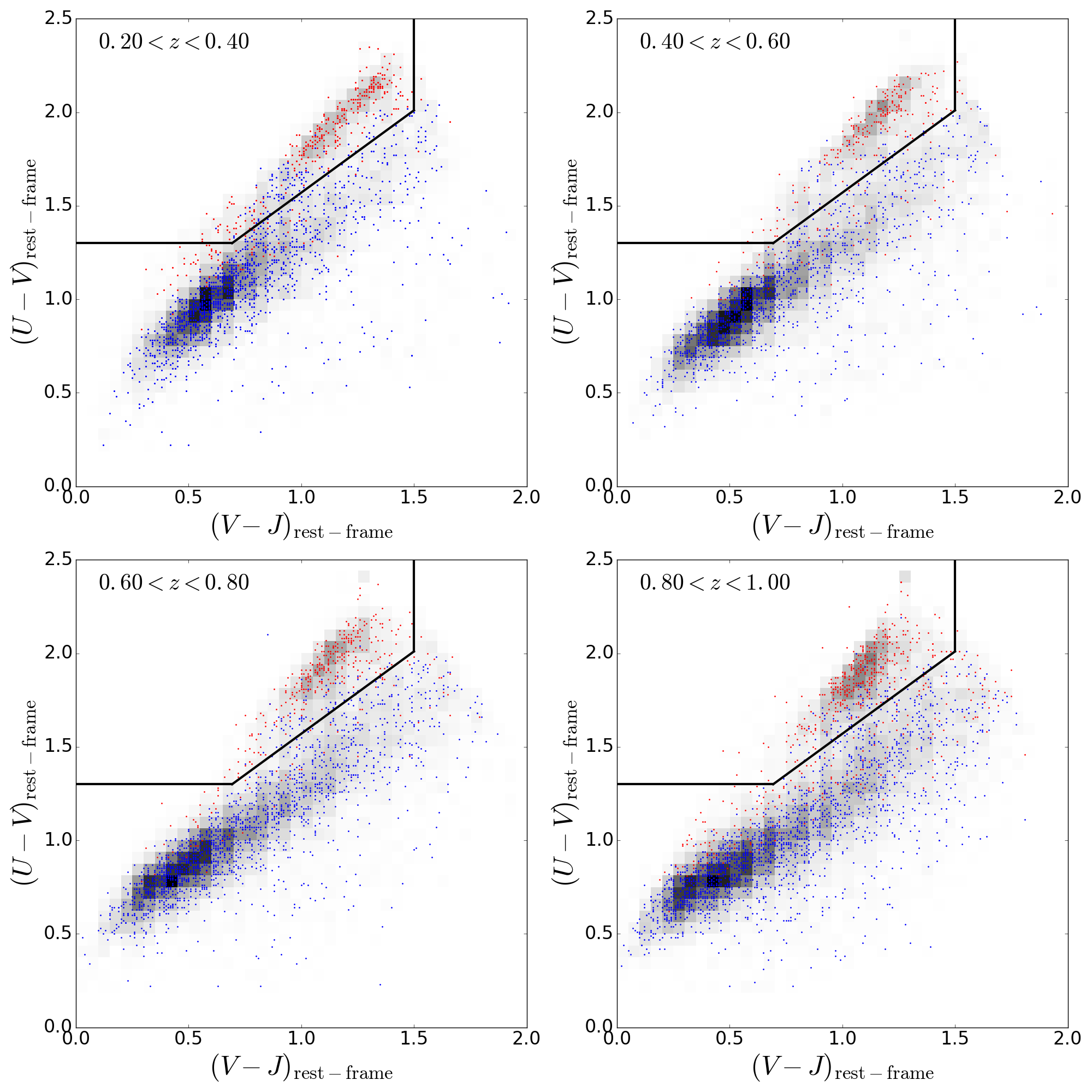}
    \caption{Color-color diagrams for COSMOS/UltraVISTA galaxies binned by redshift. Black lines divide the the galaxies into quiescent (upper region) and star-forming (lower region) population. A subsample of the galaxies are plotted as points that are colored red (quiescent, $\mathrm{sSFR} < 10^{-11} \, \mathrm{yr}^{-1}$) or blue (star-forming, $\mathrm{sSFR} > 10^{-11} \, \mathrm{yr}^{-1}$) based on the COSMOS/UltraVista estimated sSFR. We observe acceptable agreement between the quiescent/star-forming separation using color information and estimated sSFR.}
\label{fig:color-cosmos}
\end{figure*}

\subsection{Matching {\sc{UniverseMachine}} and COSMOS/UltraVISTA galaxies}
\label{subsec:matching}

To obtain insight into the form of the DES source galaxy HOD we match COSMOS/UltraVISTA photometry and assign intrinsic galaxy sizes to {\sc{UniverseMachine}} galaxies. In the case of both COSMOS/UltraVISTA and {\sc{UniverseMachine}} we rely on splitting galaxies into star-forming and quiescent populations. In {\sc{UniverseMachine}} this is accomplished by applying a cut at $sSFR = 10^{-11} \, \mathrm{yr}^{-1}$ as shown in Fig. \ref{fig:sSFR-dist}. For the COSMOS/UltraVISTA galaxies we rely on the well-known bimodality in color-color (rest frame $U-V$ vs. $V-J$) space to differentiate between star-forming and quiescent galaxies. Figure \ref{fig:color-cosmos} plots $U-V$ versus $V-J$ for COSMOS/UltraVISTA galaxies in four redshift bins. Points represent a subsample of the galaxies in each bin colored red (quiescent) or blue (star-forming) based on sSFR calculated from the SFRs and stellar masses reported by \citet{Muzzin_et_al_2013}. We observe a clear bimodality in all redshift bins with low and high sSFR galaxies cleanly separated in color-color space. We also plot solid black lines \citep{Muzzin_et_al_2013b} that we use to tag galaxies as either quiescent or star-forming. Quiescent galaxies are defined as,
\begin{align}
U-V &> 1.3\; \text{and} \; V-J<1.5, \; \text{for all} \; z, \\
U-V &> 0.88(V-J) + 0.69,\;  \text{for} \; 0.0 < z < 1.0, \\
U-V &> 0.88(V-J) + 0.59,\;  \text{for} \; 1.0 < z < 3.0,
\end{align}
with star-forming galaxies making up the remaining population. These cuts were originally defined by \citet{Williams_et_al_2009} to maximize the difference in specific star formation rates between the two populations and have since been adjusted by \citet{Muzzin_et_al_2013} to account for the difference in UltraVISTA rest-frame color distributions.

To match COSMOS/UltraVISTA photometry to a redshift snapshot of {\sc{UniverseMachine}} galaxies we use the following procedure:
\begin{enumerate}
    \item Select COSMOS/UltraVISTA galaxies in the range $z_\mathrm{UM} - 0.05 < z_\mathrm{phot} < z_\mathrm{UM} + 0.05$, where $z_\mathrm{UM}$ is the redshift of the relevant {\sc{UniverseMachine}} catalog and $z_\mathrm{phot}$ is the photometrically estimated redshift for each COSMOS/UltraVISTA galaxy. This binning is coarse relative to the photometric redshift uncertainty of our COSMOS/UltraVISTA galaxies. This should mitigate the impact of photometric redshift uncertainties, and as we will see in later sections the resulting HODs do not evolve significantly within a redshift interval comparable to the photometric redshift uncertainty.
    \item Separate mock and data galaxies into quiescent and star-forming populations as described above. In the case of {\sc{UniverseMachine}} this is done directly with the $\mathrm{sSFR}$. In the case of COSMOS/UltraVISTA galaxies this separation is determined in rest-frame color-color space.
    \item Finally for each {\sc{UniverseMachine}} galaxy we assign the photometry (i.e. $i$-, $r$-, $z$-band magnitudes) of the COSMOS/UltraVISTA galaxy that is closest in stellar mass and of the same population (quiescent/star-forming).
\end{enumerate}

\begin{figure}
\centering \includegraphics[width=0.45\textwidth]{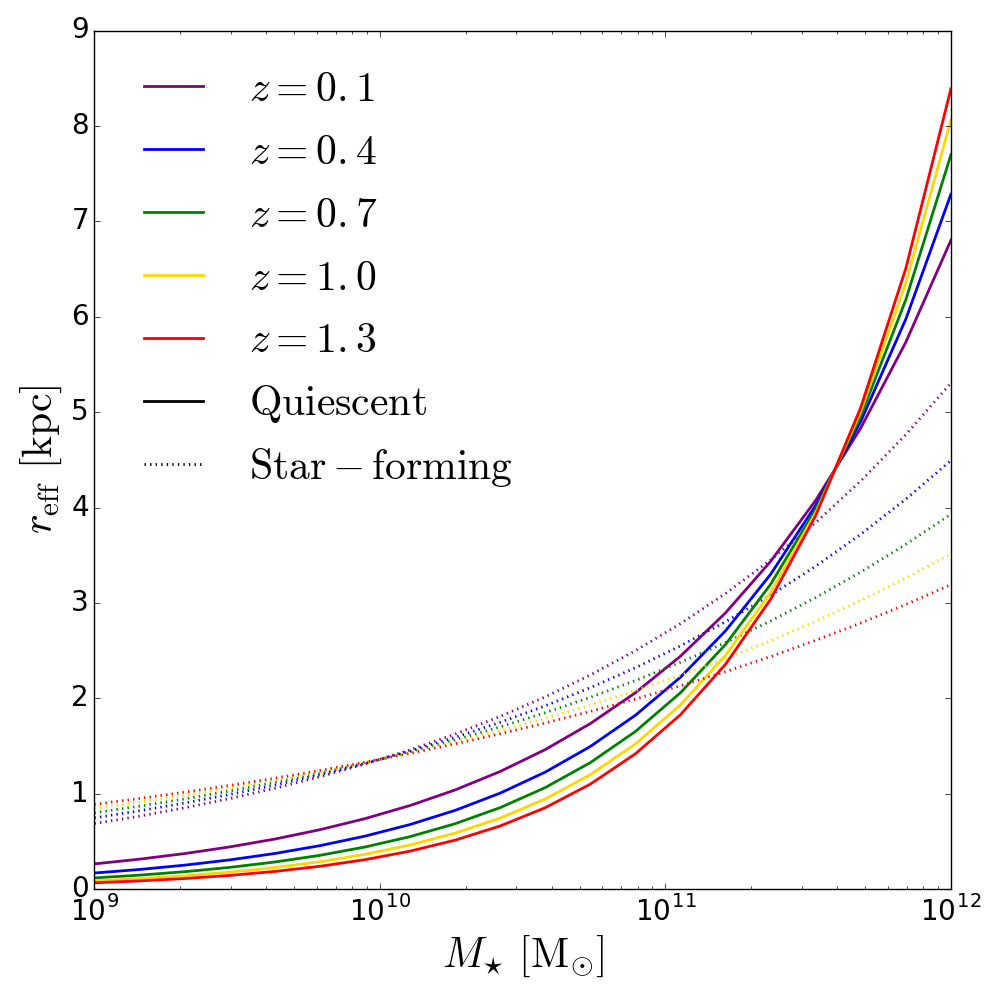}
    \caption{Redshift dependent size-mass relations from \citet{Mowla_et_al_2019a}. There are separate relations for quiescent (solid-line) and star-forming (dashed-line) galaxies that evolve significantly with redshift, and cross over each other at high stellar mass.}
\label{fig:size-mass}
\end{figure}

We also assign sizes using the size-stellar mass relations of \citet{Mowla_et_al_2019a}\footnote{See also \citet{vdWel_et_al_2014, Mowla_et_al_2019b}} for galaxies $z < 3$ calibrated from COSMOS-DASH and 3D-HST/CANDELS data. \citet{Mowla_et_al_2019a} provide separate redshift-dependent relations for star-forming and quiescent galaxies of the form
\begin{align}
    r_\mathrm{eff}(m_\star) / \mathrm{kpc} &= A \times m_\star^\alpha, \\
    m_\star &= M_\star / (7 \times 10^{10} M_\odot),
\end{align}
where $r_\mathrm{eff}$ is effective radius and $M_\star$ is the galaxy stellar mass, the semi-major axis of the ellipse containing half the flux of the galaxy, and
\begin{align}
    \log A &= \begin{ecases*}
    -0.29 \log \left( 1 + z \right) + 0.91 & star-forming, \\
    -0.54 \log \left( 1 + z \right) + 0.72 & quiescent,
    \end{ecases*}\\
    \alpha &=  \begin{ecases*}
    -0.15 \log \left( 1 + z \right) + 0.31& star-forming, \\
    0.31 \log  \left( 1 + z \right) + 0.44 & quiescent.
    \end{ecases*}
\end{align}
Figure \ref{fig:size-mass} shows these relations over a range of redshifts. Both quiescent and star-forming galaxies increase in size with increasing stellar mass. The relation of size and redshift is non-monotonic reversing for  high-mass ($M_\star>4\times10^{11}\,\Msun$) quiescent and low-mass  ($M_\star<10^{10}\,\Msun$) star-forming galaxies. At low stellar masses star-forming galaxies tend to be larger than their quiescent counterparts. This trend reverses between $M_\star \approx 1${-}$2\times10^{11}\,\Msun$ depending on galaxy redshift. 

In addition to the mean of these relations we also consider the scatter in $r_\mathrm{eff}$ at fixed stellar mass denoted $\siglnReff$. \citet{Mowla_et_al_2019b} report this scatter to be roughly log-normal and constant in stellar mass with values ranging between 0.2 and 0.3 dex for both quiescent and star-forming galaxies. They also report minimal redshift evolution of this quantity. These results are in good agreement with others from the literature \citep{Ichikawa_et_al_2012, vdWel_et_al_2014}. Therefore we will adopt a constant lognormal scatter with respect to stellar mass and redshift with value $\siglnReff = 0.25$ as our fiducial model of size-mass scatter.

To apply the source galaxy selection criteria we must convert $r_\mathrm{eff}$ into the size $T$ (described in detail in Section \ref{sec:source_cuts}). Assuming a circular two-dimensional Gaussian intensity profile for source galaxies and relating its second moments to the half-light radius $r_\mathrm{eff}$ yields,
\beq
T = \frac{r_\mathrm{eff}^2}{\left[\mathrm{erf}^{-1}\left(\frac{1}{\sqrt{2}}\right)\right]^2},
\eeq
which we convert into angular units using the redshift of each of our simulation snapshots. For our star-forming galaxies we additionally randomly incline them with respect to the line-of-sight when computing their apparent size.

\section{Modeling the HOD of DES source galaxies}
\label{sec:HOD}
\subsection{Applying source galaxy selection to mock galaxies}
\label{subsec:cuts}

\begin{figure*}
\centering \includegraphics[width=1.0\textwidth]{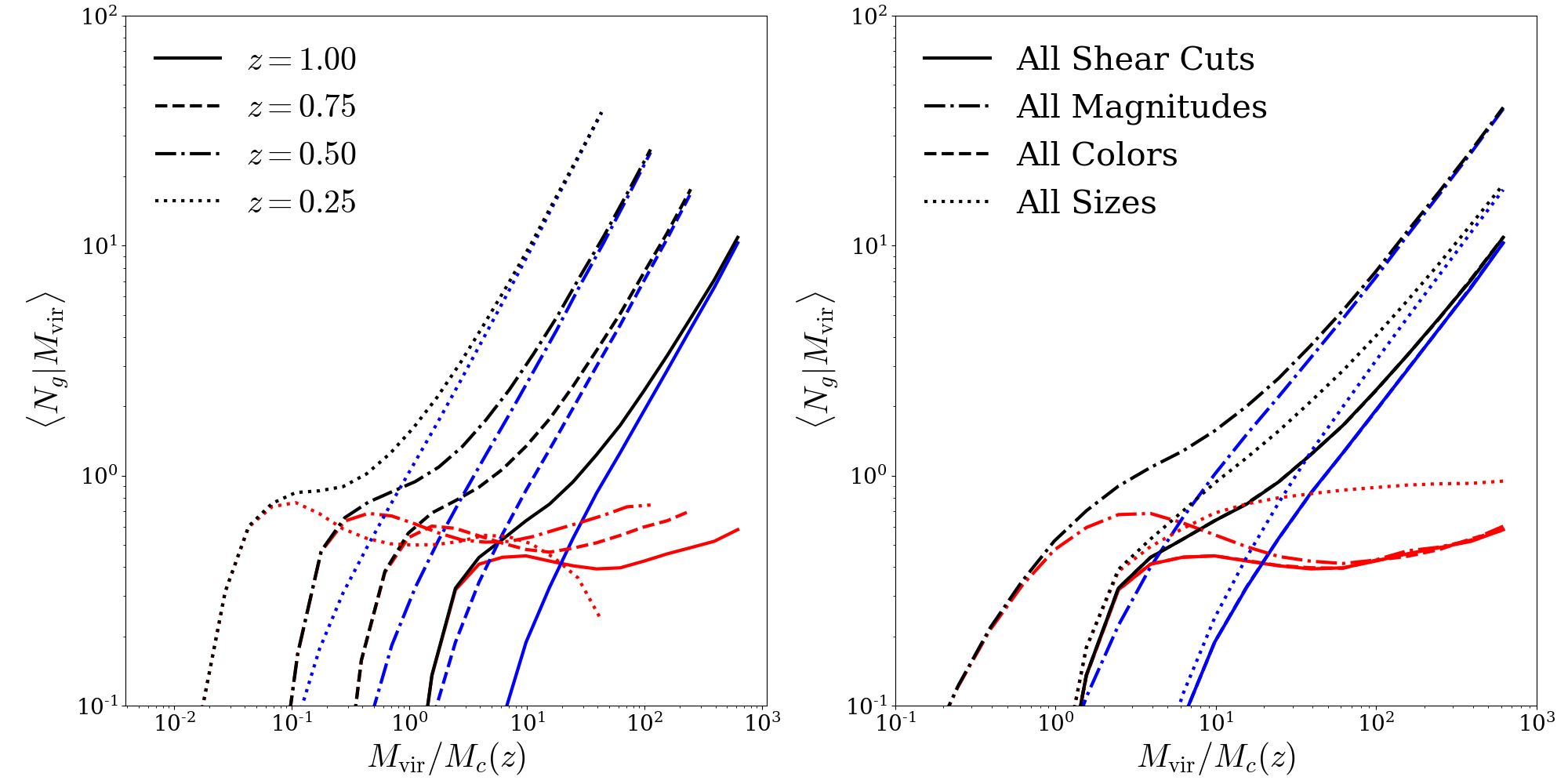}
    \caption{The HOD of {\sc{UniverseMachine}}-COSMOS/UltraVISTA matched source galaxies for central (red), satellite (blue) and all (black) galaxies. The left-hand panel shows how the source HOD varies with redshift. We observe significant redshift evolution in the range $z = 0.25 - 1.00$, particularly in the characterstic mass scales to host centrals and satellites. The right-hand panel shows the effect of relaxing selection criteria at $z = 1.00$. The solid line shows the HOD obtained with all selection criteria applied, the dot-dashed curves show the impact of relaxing the magnitude cuts, the dashed shows the impact of relaxing color cuts, and the dotted shows the impact of relaxing size cuts. We see that the significant central-incompleteness we observe is driven by cuts on galaxy size (see text). }
\label{fig:full_HOD_results}
\end{figure*}

With COSMOS/UltraVISTA photometry matched to {\sc{UniverseMachine}} galaxies and sizes assigned we can apply the DES-Y3 source selection criteria to our mock galaxies and directly measure their HOD. The selection criteria are described in detail in Section \ref{sec:source_cuts} and summarized here,
\begin{align}
18 < \, &i < 23.5, \label{eq:mag1} \\
15 < \, &r < 26, \label{eq:mag2} \\
15 < \, &z < 26, \label{eq:mag3} \\
-1.5 < \, &r - i < 4, \label{eq:color1} \\
-1.5 < \, &z-i < 4, \label{eq:color2} \\
\frac{T_\mathrm{PSF}}{2} < \, &T < 10 \, \mathrm{arcsec}^2, \label{eq:size}
\end{align}
where $T_\mathrm{PSF} = 0.33 \, \mathrm{arcsec}^2$ is the average for DESY3 data source galaxies. The left-hand panel of Fig. \ref{fig:full_HOD_results} shows the predicted source galaxy HOD over a range of redshifts for centrals (red), satellites (blue), and all galaxies (black). 

We plot the mass-dependence of the HOD in terms of the characteristic collapse mass $\Mstar$ satisfying $\sigma(\Mstar(z)) = \delta_c$ where $\sigma(M,z)$ is the square root of the variance in the linear density field and $\delta_c = 1.686$ is the threshold linear overdensity for spherical collapse. Across redshifts we see that the central occupation is roughly consistent with the standard form, i.e. an error-function, albeit with significant incompleteness. 

We find that the central occupation does not converge to a single value with increasing mass at any redshift, though it is difficult to distinguish between genuine mass-dependent behavior and noise (from finite simulation box size). In either case the contribution to the galaxy bias from high mass halos will be dominated by the satellite term.  We also see that the level of central incompleteness is dependent on redshift, roughly $60 - 70\%$ halos host a central at $z = 0.50$, with this fraction decreasing to roughly $40\%$ at $z = 1.0$. However, this behavior is not monotonic across the entire redshift range we consider, at $z = 0.25$ we see a precipitous decline in high-mass occupation driven by the large size cut, $T < 10 \, \mathrm{arcsec}^2$.

In the case of the satellites we see that at all redshifts the satellite occupation takes the familiar form of a power-law with an exponential cut-off at low masses. We observe that the cut-off mass and the characteristic mass to host a satellite depend on redshift and that both increase with increasing redshift. In contrast the slope of the satellite-occupation power law only evolves negligibly with redshift.

In the right-hand panel of Fig. \ref{fig:full_HOD_results} we investigate the impact of the selection criteria at $z = 1.0$. Dot-dashed curves show the impact of relaxing our magnitude cuts (equations \ref{eq:mag1}{-}\ref{eq:mag3}), dashed curves show the impact of relaxing the color cuts (equations \ref{eq:color1} and \ref{eq:color2}), and dotted curves show the impact of relaxing the size cuts (equation \ref{eq:size}). We see that both our magnitude and size cuts significantly change the normalization of the satellite occupation but have less of an effect on the slope. The color cuts have negligible impact on the satellite occupation, indicating that they are redundant with the magnitude and size cuts.

The impact on the central term is more complicated. We see that the degree and form of the observed incompleteness is driven by our size cut, i.e. when the size cut is relaxed we obtain a standard error-function form for our central occupation. This effect is ultimately a result of a complex interplay between the stellar-to-halo mass relation, the quenched fraction and the size-stellar mass relations for star-forming and quiescent galaxies. In Fig. \ref{fig:size-mass} we see that the quiescent size-mass relation crosses over the star-forming relation at large stellar mass. Similarly the quenched fraction evolves with stellar/halo mass. The redshift evolution of this behavior observed in the right-hand panel of Fig. \ref{fig:full_HOD_results} becomes more complicated as the large and small size cuts are fixed in angular size and therefore correspond to different physical sizes at different redshifts.

At the low mass end we see that the occupation is most affected by our magnitude cuts, which remove low magnitude galaxies hosted by low mass halos. When the magnitude cuts are relaxed the characteristic mass to host a central decreases by almost an entire dex. Additionally we see that the magnitude cuts largely determine the scatter in the characteristic mass to host a central. When these are relaxed the width in the transition from zero occupation is broadened significantly. As in the case of the satellite occupation we see that the color cuts are almost entirely redundant with the size and magnitude cuts.

\subsection{Analytic HOD parameterization}
\label{subsec:HOD_parametrization}

In Fig. \ref{fig:full_HOD_results} we have shown the HOD of mock source galaxies in our {\sc{UniverseMachine}}-COSMOS/UltraVISTA matched catalog. Across a wide redshift range we observe a power-law satellite term consistent with a standard HOD parameterization. The central term exhibits significant incompleteness with complex mass-dependence. Fortunately, the more complicated features of this incompleteness occur at high mass where the central contribution to the galaxy correlation function is small relative to that from the satellite term. Therefore we adopt a modified version of the standard HOD parameterization \citep[e.g.][]{Berlind_2002, Zheng_et_al_2005} for the source central and satellite occupations as a fully analytic option for modeling the DES-Y3 source sample HOD,
\begin{align}
\langle N_\mathrm{cen} | M_h \rangle &= \frac{\fcen(M)}{2} \left[1 + \mathrm{erf}\left( \frac{\log M - \log \Mmin}{\siglogM} \right) \right] \\
\langle N_\mathrm{sat} | M_h \rangle &= \frac{\langle N_\mathrm{cen} | M_h \rangle}{\fcen(M)}  \left( \frac{M}{M_1} \right)^\alpha \\
\fcen(M) &= \mathrm{min}\left[ \fcenO \left(1 + \frac{M}{10^{\log \Mmin + \siglogM}} \right)^\beta, 1.0 \right]
\end{align}
where the familiar parameter $\Mmin$ represents the characteristic halo mass required to host a central ($\langle N_\mathrm{cen} | \Mmin \rangle = 0.5$), the parameter $\siglogM$ determines the sharpness of the transition from $\langle N_\mathrm{cen} \rangle = 0.0$ to $\langle N_\mathrm{cen} \rangle = 1.0$, $M_1$ is the characteristic halo mass required to host a satellite ($\langle N_\mathrm{cen} | \Mmin \rangle = 1.0$), and $\alpha$ is the slope of the satellite power law. We also include a linear mass-dependent incompleteness term characterized by $\fcen(M)$, with normalization $\fcenO$ and slope $\beta$. This is to model the nontrivial central incompleteness which we observe in Fig. \ref{fig:full_HOD_results}.

\begin{figure}
\centering \includegraphics[width=0.45\textwidth]{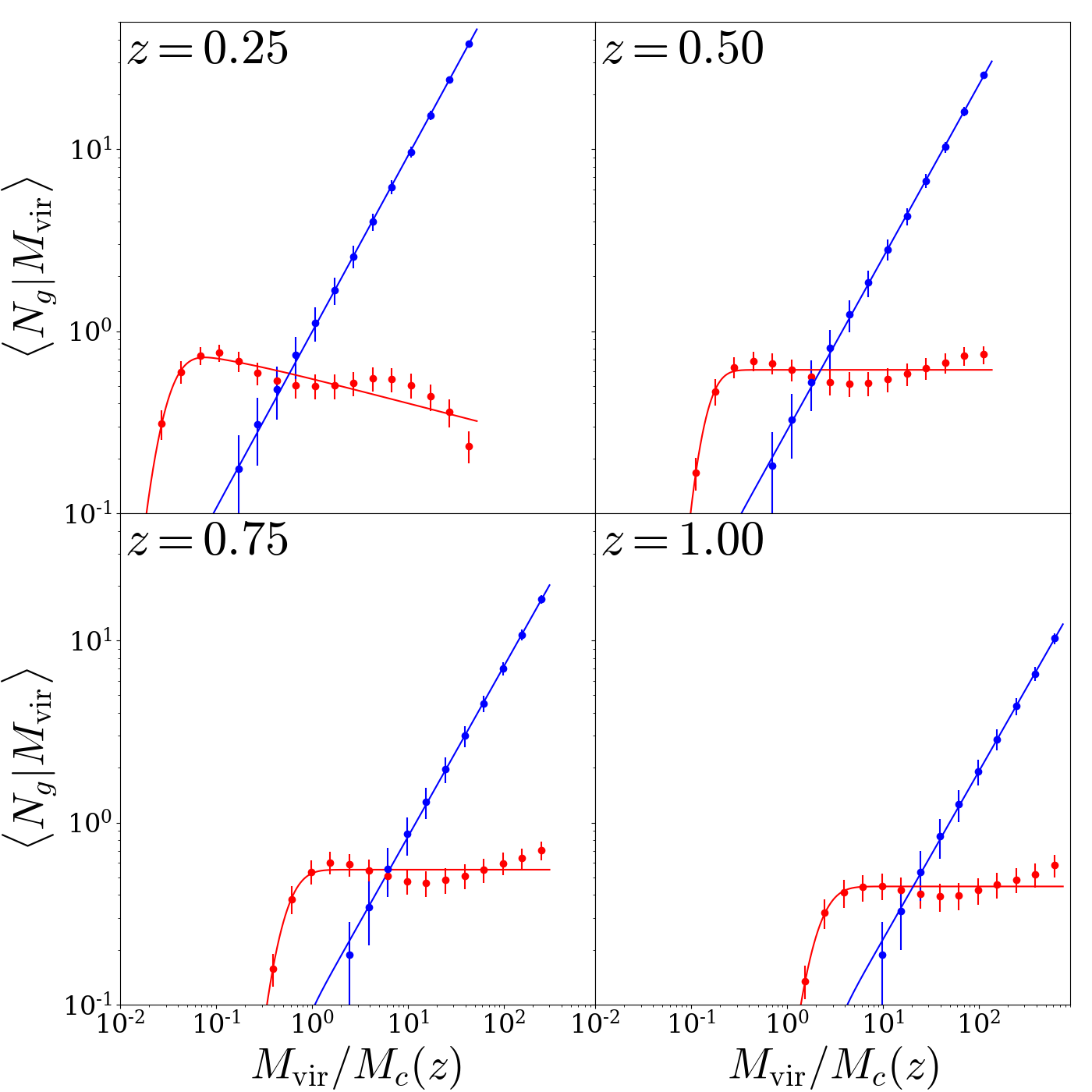}
    \caption{Best fitting HOD models for our mock source galaxies at $z = 0.25$ (top-left), $z = 0.50$ (top-right), $z = 0.75$ (bottom-left), and $z = 1.00$ (bottom right). We see that the satellite term is very well fit by a standard power law, while the modified central term captures overall changes in incompleteness as a function of mass.} 
\label{fig:HOD_fits}
\end{figure}

Figure \ref{fig:HOD_fits} shows the result of fitting this parameterization to our fiducial source HODs at redshifts $z = 0.25$, $0.50$, $0.75$ and $1.00$. This fitting is done by minimizing the $\chi^2$ for the central and satellite terms separately in each redshift bin. We see that across our redshift range the satellite occupation is well fit by a standard power-law occupation. The overall mass dependence of the observed central incompleteness is described by our parameterization. At high masses we see some degree of model mis-specification but in this mass regime the satellite occupation dominates the HOD and galaxy bias.

\section{HOD priors for cosmological analyses}
\label{sec:hodpriors}

\subsection{HOD sensitivity to underlying assumptions}
\label{sec:robust}

\begin{figure}
\centering \includegraphics[width=0.45\textwidth]{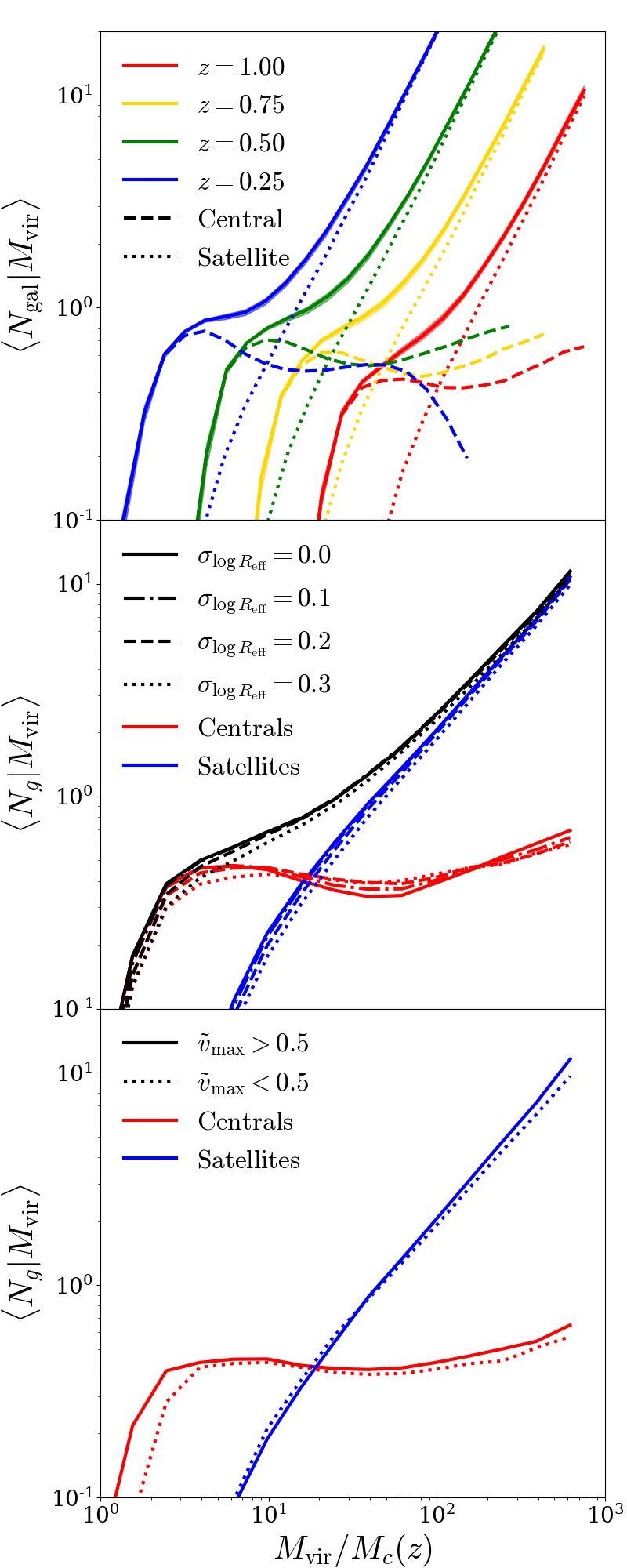}
    \caption{Robustness tests for our {\sc{UniverseMachine}}-based source HOD results described in Section \ref{sec:robust}. The top-panel shows the error on our fiducial HOD due to {\sc{UniverseMachine}} uncertainties across the redshift range we consider. The middle-panel shows the impact of the variations in the assumed scatter in the size-mass relation. Finally the bottom-panel shows the level of galaxy assembly bias we observe in our mocks.}
\label{fig:HOD_robust}
\end{figure}

Thus far we have obtained results for the DES-Y3 source HOD assuming a single realization the {\sc{UniverseMachine}} algorithm, a single fiducial value for the scatter in the size-mass relation, and we have ignored the potential for the source sample to exhibit galaxy assembly bias. In Fig. \ref{fig:HOD_robust} we examine the impact of each of these assumptions on our results. 

Beginning with the top-panel we test the impact of {\sc{UniverseMachine}} uncertainties on our HOD results. The {\sc{UniverseMachine}} fits a variety of galaxy observational data using a 44-parameter forward modeling framework that parameterizes galaxy star formation rate as a function of host halo properties. Our matching scheme relies on the specific star formation rates and stellar masses of galaxies that this forward model predicts and is therefore sensitive to uncertainties in these model parameters. 

To quantify this we draw 100 realizations from the {\sc{UniverseMachine}} posterior and apply these model realizations to the Bolshoi-Planck simulation. For each of these 100 realizations we then recalculate the expected source HOD. The top-panel of Fig. \ref{fig:HOD_robust} shows the result of this test at redshift $z = 0.25$, $0.50$, $0.75$, and $1.00$. Solid lines show the mean total HOD occupation, while dashed lines show the central component and dotted lines show the satellite componenent. The colored bands show the $1{-}\sigma$ error from our 100 realizations. When propagated through our matching scheme the uncertainty of the HOD on the {\sc{UniverseMachine}} posterior is at the $3{-}5\%$ percent level. The resulting uncertainty on the satellite fraction is $4-5\%$. We will further quantify the level of this variation in the context of producing HOD-priors for cosmological analyses in subsequent sections. 

In the middle-panel of Fig. \ref{fig:HOD_robust} we investigate the impact of scatter in the size-mass relation on the HOD. For satellites increasing the scatter reduces the occupation at all masses. For centrals the impact of changing the scatter is mass-dependent. At low and high masses increased scatter decreases the occupation due to the small and large size cuts. Meanwhile in an intermediate mass range increased scatter increases the occupation. This mass-dependent behavior tends to flatten out the central occupation, which has the deepest local minimum at intermediate masses in the case of no scatter. The detailed features of the observed central incompleteness are due to the small and large size cuts used to select source galaxies, as the scatter increases and tends to dominate over the mean size-mass relation these features become less salient and the central occupation is flattened.

Finally in the bottom-panel of Fig. \ref{fig:HOD_robust} we investigate the level of galaxy assembly bias present in the source sample. Galaxy assembly bias refers to the possibility for galaxy occupation to depend on properties other than mass,
\beq
\langle N | M_h \rangle \neq \langle N | M_h, S \rangle,
\eeq
where $S$ is some secondary property \citep[e.g.][]{Croton_et_al_2007, Zu_et_al_2008, Zentner_Hearin_vdBosch_2014, McCarthy_et_al_2019, Zentner_et_al_2019, Salcedo_et_al_2022b, Wang_et_al_2022}. If there also exists a halo assembly bias \citep[e.g.][]{Sheth_Tormen_2004,Gao_2005,Harker_et_al_2006, Wechsler_2006,
Gao_White_2007, Jing2007, Wang_Mo_Jing_2007, Li_Mo_Gao_2008, Faltenbacher_White_2010,
Mao_Zentner_Wechsler_2018, Salcedo_2018, Sato-Polito_et_al_2019, Xu_Zheng_2018, Johnson_et_al_2019} signal with respect to the same property then the galaxy bias will differ from that predicted by a standard HOD model.  

In the case of both centrals and satellites we compute the mean occupation for haloes with maximum circular velocity $v_\mathrm{max}$ greater (solid) and less (dotted) than the median for haloes of that mass. We see that the haloes with high $v_\mathrm{max}$ for their mass are more likely to host central galaxies, particularly at low masses. In contrast the satellite occupation is not significantly affected by $v_\mathrm{max}$. This is a clear indication of galaxy assembly bias in our mock source galaxy sample that must be properly marginalized over to produce robust cosmological constraints. Though we leave determining how exactly this will be accomplished to further work, we note that a parametric correction to the galaxy bias as a function of assembly bias strength motivated by our results is likely to be sufficient.

\begin{table*}
   \centering
   \caption{Mean best fitting HOD parameters and dispersion from the 1000 {\sc{UniverseMachine}}, scatter and bias realizations.}
    \begin{tabular}{ccccccc}      
    \hline
    $z$ & $\siglogM$ & $\log \left( \Mmin/\Mstar \right)$ & $\log \left(M_1 / \Mstar\right)$ & $\alpha$ & $\fcenO$ & $\beta$ \\
    \hline
0.25 & $0.41\pm0.03$ & $-1.51\pm0.03$ & $-0.02\pm0.03$ & $0.96\pm0.01$ & $0.80\pm0.05$ & $-0.16\pm0.02$\\
0.50 & $0.26\pm0.04$ & $-0.87\pm0.04$ & $0.55\pm0.03$ & $0.95\pm0.02$ & $0.65\pm0.06$ & $0.03\pm0.13$\\
0.75 & $0.25\pm0.02$ & $-0.31\pm0.03$ & $1.10\pm0.04$ & $0.94\pm0.03$ & $0.57\pm0.04$ & $0.02\pm0.11$\\
1.00 & $0.28\pm0.05$ & $0.31\pm0.05$ & $1.73\pm0.04$ & $0.95\pm0.04$ & $0.48\pm0.03$ & $0.03\pm0.24$\\
1.25 & $0.38\pm0.04$ & $1.03\pm0.10$ & $2.74\pm0.23$ & $1.01\pm0.18$ & $0.23\pm0.04$ & $-0.07\pm0.19$\\
    \hline
   \end{tabular}
\label{tab:HOD-best}
\end{table*}

\subsection{Analytic HOD priors for cosmological analysis}
\label{subsec:prior_real}

By matching COSMOS/UltraVISTA photometry to {\sc{UniverseMachine}} mock galaxies we have constructed mock galaxy catalogs with the properties necessary to apply DES-Y3 source galaxy selection criteria in order to study their galaxy-halo connection and develop an HOD model for future cosmological analyses. As discussed in Section \ref{sec:robust} our best-fit HOD is sensitive to variations in the underlying parameters. Our framework allows us to quantify these uncertainties to obtain priors that can be used in future cosmological analyses. 

To produce such priors we must propagate the systematic uncertainties in our matching scheme to uncertainties on HOD parameters. The main sources of systematic uncertainties in this process are,
\begin{itemize}
\item Uncertainty in the relation between COSMOS/UltraVISTA and DES photometry. 
\item Posterior uncertainties on the {\sc{UniverseMachine}} model parameters  
\item Uncertainties in the galaxy size-stellar mass relation.
\end{itemize}
In each case we can make reasonable assumptions about the level of uncertainty, and therefore we can convert these assumptions into priors on HOD parameters.

To do this we generate noisy and biased realizations of our fiducial DES source selection. We use the 100 draws from the {\sc{UniverseMachine}} posterior applied to the Bolshoi-Planck simulation described in Section \ref{sec:robust}. We additionally treat the bias and gaussian scatter in each of  $T$, $i$, $r$, and $z$ as parameters.  For these 8 parameters we Latin-hypercube sample over conservative ranges informed by the expected errors in DES photometry and range in galaxy size-stellar mass relations reported in the literature.  To each {\sc{UniverseMachine}} realization we assign 10 sets of bias and scatter parameters and recompute the source HOD for each.  This process generates a total of 1000 realizations of the DES source HOD at each of our redshift snapshots.

The variation in resulting HOD relations at a given redshift is therefore representative of our uncertainty on the underlying quantities that impact DES source selection. We fit the central and satellite term of each of these HOD realizations with the parametrization presented in Section \ref{subsec:HOD_parametrization}. The mean and standard deviation of the fitted HOD parameters are reported in Table \ref{tab:HOD-best} and will serve as the basis for priors in a future HOD analysis of the DES source galaxies. We note that the redshift evolution of the HOD we observe in Table \ref{tab:HOD-best} is not significant in redshift intervals equal to the photometric uncertainty of the COSMOS/UltraVISTA galaxies.

\subsection{Non-parametric emulation of the HOD}

An alternative method to model the source HOD and specify priors for a future analysis is to non-parametrically emulate the HOD in terms of the uncertainties described above. Priors in these uncertainties can then be converted into priors on the overall source HOD, and by non-parametrically emulating the HOD we reduce the impact of model mis-specification. Because the {\sc{UniverseMachine}} has 44 model parameters it is difficult to emulate the entire parameter space, and additionally it is not clear how to set priors on each of these parameters. The most important quantity for our matching scheme that {\sc{UniverseMachine}} predicts is the stellar mass function, and therefore we will focus on the deviation from the fiducial stellar mass function as our main criterion.

We generate a set of realizations using a similar process to that described above. Instead of generating full {\sc{UniverseMachine}} realizations we treat the deviation from the fiducial stellar mass function in units of posterior standard deviation $\sigma$ as a parameter that we Latin-hypercube sample in addition to the scatter and bias in $T$, $i$, $r$ and $z$. We generate 1000 such parameter samples and assign 50 to each of 20 redshift snapshots in the range $z = 0.10 - 1.26$. For each of the 1000 mock realizations we then compute the source HOD.

To emulate this HOD in terms of our parameters and redshift we perform a Gaussian process regression with a squared-exponential kernel in each mass bin of $\langle N_\mathrm{cen} | M_\mathrm{vir} \rangle$ and $\langle N_\mathrm{sat} | M_\mathrm{vir} \rangle$. The advantage of this procedure is two-fold, 1) it does not assume an analytic form of the HOD or its redshift evolution and therefore can capture the complex behavior we observe in Fig. \ref{fig:full_HOD_results} and 2) it connects the HOD to observational properties that we can directly place informative priors on. Note that we emulate in bins of $M_\mathrm{vir}$ rather than $M_\mathrm{vir} / \Mstar$.

\begin{figure}
\centering \includegraphics[width=0.45\textwidth]{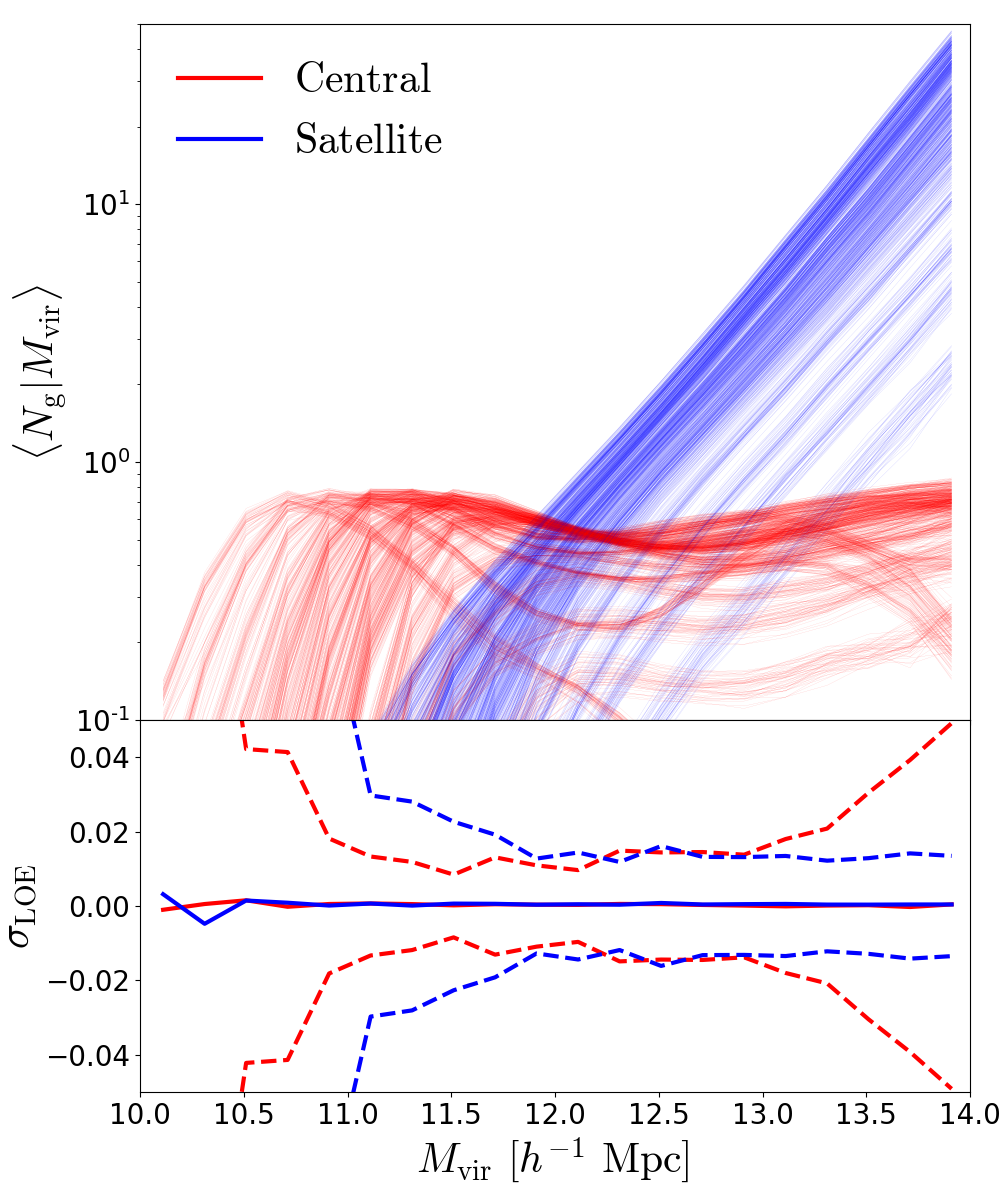}
    \caption{Training samples (top-panel) and leave-one-out emulation errors and biases (bottom-panel) for the central (red) and satellite (blue) contributions to our source HOD. The top panel plots the central and satellite term for each of our 1000 training samples drawn from a wide range of redshifts and observational error models. The bottom panel shows the leave-one-out $1\sigma$ error (dashed) and bias (solid) of our emulation as a function of host halo mass.} 
\label{fig:emu_acc}
\end{figure}

Figure \ref{fig:emu_acc} shows the results of this emulation. In the top panel we show the set of HOD realizations used to train our emulator split into central (red) and satellite (contributions). In the bottom panel we plot the $1\sigma$ leave-one-out error and the bias in our emulation as a function of mass for our central and satellite data. In both cases we see that over the relevant range in halo mass we achieve $1-3\%$ errors and are unbiased. In a future cosmological analysis this emulator can be used to directly model and constrain the source HOD.

\subsection{Robustness to cosmology}

The results presented thus far have been based on a single cosmology used to generate the SMDPL and Bolshoi-Planck simulations. This raises two questions, whether the form of the source HOD depends on cosmology and the extent to which the priors on the HOD we have derived are cosmology dependent. In an HOD context the key quantities to predict galaxy summary statistics are the halo mass function $\frac{dn}{dM_h}$, the halo bias $b_h$, and the HOD itself $\langle N_g | M_h \rangle$. For example the large scale galaxy bias is given by
\beq
b_g = \frac{1}{n_g} \int d M_h \frac{d n}{d M_h} b(M_h) \langle N_g | M_h \rangle,
\eeq
which is an HOD and halo-mass function weighted sum of the halo bias where $n_g$ is the galaxy number density. Therefore cosmology can affect galaxy formation solely through the halo mass function or halo bias, each of which are easily modeled given current prescriptions. This introduces degeneracies between cosmological and HOD parameters in an analysis of galaxy clustering without any dependence of the HOD on cosmology.

More complicated is the possibility for cosmology to impact the galaxy HOD itself. Previous literature on the cosmological dependence of galaxy formation has not typically separated these effects. To put the question differently, for a given halo mass at a given redshift how do the galaxy populations formed in such halos depend on cosmology?

Galaxy formation is the complicated product of a variety of physical processes \citep[e.g.][]{Somerville_Dave_2015} many of which are not yet fully understood and not resolvable in cosmological hydrodynamic simulations. In the standard picture, every galaxy is formed within a halo with initial properties associated with its host's mass and accretion history. In addition to forming a central galaxy at their center, halos acquire new galaxies through mergers with other halos. The orbits of the newly acquired satellites will gradually decay and lead to interactions and mergers with the central. The standard HOD formalism assumes that the outcome of this linked halo and galaxy formation process depends only on host halo mass. If this assumption holds then the key ingredients to predict the HOD at a given redshift are the stellar-to-halo mass relation (SHMR) and the subhalo mass function. The degeneracy between redshift and cosmological parameters is mostly accounted for by considering the HOD in units of the redshift- and cosmology-dependent typical collapse mass $\Mstar$.

The baryon density $\Omega_b$ will naturally impact galaxy formation and the resulting SHMR. The baryon fraction in halos is expected to be roughly 90\% of the universal value $\Omega_b / \Omega_m$ and to be fairly independent of redshift up to $z = 1$, and halo mass above $\log M_h \approx 10.5$ \citep{Crain_et_al_2007}. Given this independence we would expect a change in $\Omega_b$ at fixed $\Omega_m$ to change the normalization of the central and satellite SHMR. This effect will enter our formalism through shifts in $M_\mathrm{min}$, $M_1$, and $f_\mathrm{cen,0}$. The parameter $\siglogM$ is tied to the scatter in the SHMR and is therefore not effected by shifts in it's normalization. In the top right panel of Fig. \ref{fig:full_HOD_results} we see that $M_\mathrm{min}$ is set by the magnitude cuts, $f_\mathrm{cen,0}$ is set by the size cuts and $M_1$ is controlled by both magnitude and size cuts. By the same logic we would expect shifts in $\Omega_m$ at fixed $\Omega_b$ to produce shifts in $M_\mathrm{min}$, $M_1$ and $f_\mathrm{cen,0}$. 

The key halo-model ingredient to predict the satellite occupation is the subhalo-mass-function (SHMF) and satellite SHMR. As previously mentioned shifts in $\Omega_m$ and $\Omega_b$ are expected to impact the normalization of the satellite SHMR, and therefore $M_1$. Changes in the shape of the SHMF could potentially impact $\alpha$. Recently \citet{Ragagnin_et_al_2023} investigated the cosmological dependence of the satellite galaxy abundance in the {\it{Magneticum}} hydrodynamical simulations. They find that the normalization of the satellite abundance function, i.e. $M_1$, depends weakly on cosmology parameters parcticularly $\Omega_m$ and $\Omega_b$. They also find that the logarithmic slope of the satellite occupation is insensitive to cosmology. This would suggest that we can safely treat the cosmological dependence of $\alpha$ as negligible. These arguments suggest that the functional form of our source HOD is insensitive to cosmology.

In this discussion we have assumed that the HOD depends only on mass and can therefore be predicted with knowledge of the SHMR and SHMF, implicitly assuming a lack of galaxy assembly bias. Fortunately this is a safe assumption as \citet{Contreras_et_al_2021} recently investigated the cosmological dependence of halo and galaxy assembly bias and found it to be ``practically negligible.'' This suggests that predictions of the level of galaxy assembly bias for DES-Y3 source galaxies from our methodology will not be sensitive to our assumption of a given cosmology, and therefore prior information on galaxy assembly bias from this study can reliably be incorporated into future cosmological analyses.

\section{Conclusions}
\label{sec:conclusions}

In this paper we have studied the galaxy-halo connection of DES-Y3 source galaxies in order to enable a future ``lens equal source'' analysis that includes small scales. We have developed a novel matching technique that takes advantage of the tight relation between the stellar mass and photometry of quenched and star-forming galaxies. This technique is applicable to any galaxy sample selected with photometry included in the 30 bands of COSMOS/UltraVISTA data. 

Our study has revealed a complex phenomenology in the DES-Y3 source galaxy-halo connection that is not captured by standard HOD forms. In particular, our mocks indicate that DES-Y3 source galaxies suffer from significant central incompleteness, primarily driven by the size cuts that define the source sample. Additionally, we also observe this incompleteness to be strongly redshift and mass dependent. Indeed, we observe significant redshift evolution of both the central and satellite occupations of source galaxies over the redshift range $z = 0.1{-}1.25$. 

We also observe the presence of central galaxy assembly bias in our mock source galaxies, with high-$v_\mathrm{max}$ halos being more likely to host source centrals than their low-$v_\mathrm{max}$ counterparts at fixed mass. To describe this behavior, we have developed an analytic form for the source HOD. We have shown that this analytic HOD form can describe the complex behavior of the source HOD across a wide range of redshifts, in particular the non-trivial central incompleteness introduced by source selection on galaxy sizes.

This halo-model approach allows us to convert our  knowledge of galaxy formation and observable errors in DES photometry into priors on the HOD of source galaxies that can be used in future analyses. To define these HOD priors we have generated different realizations of our source HOD assuming realistic variations in photometry and the galaxy size-mass relation, and we marginalize over the {\sc{UniverseMachine}} posterior. We fit each of these realizations using our new analytic form, confirming its suitability for modeling the source HOD. The dispersion in these best fit parameters then represents our uncertainty on the source sample's galaxy-halo connection. 

We have also used this set of realizations to train an emulator that models the dependence of the source HOD on redshift, errors in photometry and sizes as well as {\sc{UniverseMachine}}'s uncertainty on the stellar mass function. This emulator describes this behavior at the 1-3\% level across the relevant mass and redshift ranges. This emulator allows us to convert prior knowledge on DES observational uncertainties into implicit priors on the source HOD. In the future, this kind of forward modeling approach can be used to integrate HOD models into likelihood analyses directly, independent of any functional form.

The principal challenge when trying to utilize information on small scales in clustering and galaxy-galaxy lensing is that the potential information gain is prohibited by model uncertainty. This model uncertainty does not only originate from a sufficiently flexible paramterization, but also from the lack of priors on the relevant parameters. In this work we have studied the halo-galaxy connection of DES-Y3 source galaxies and developed techniques to set robust and informative priors on their HOD. This halo-model approach enables us to push to small scales and allows us to incorporate prior knowledge on galaxy formation physics to constrain nuisance parameters associated with the galaxy-halo connection. These techniques represent a powerful path towards maximizing the information gain from current and future galaxy surveys and will allow us to take advantage of future developments in our understanding of galaxy formation in a cosmological context.

\section*{Acknowledgements}
We thank Andrew Hearin, Matthew Becker, ChangHoon Hahn, David Weinberg and Elisabeth Krause for valuable conversations on this work. AS, and TE are supported by the Department of Energy grant DE-SC0020215. Simulations in this paper use High Performance Computing (HPC) resources supported by the University of Arizona TRIF, UITS, and RDI and maintained by the UA Research Technologies department. Simulations were analyzed in part on computational resources of the Ohio Supercomputer Center \citep{OhioSupercomputerCenter1987}, with resources supported in part by the Center for Cosmology and AstroParticle Physics at the Ohio State University. We gratefully acknowledge the use of the {\sc{matplotlib}} software package \citep{Hunter_2007} and the GNU Scientific library \citep{GSL_2009}. This research has made use of the SAO/NASA Astrophysics Data System. 

\bibliography{masterbib2}

@Preamble{ " \newcommand{\noop}[1]{} " }

@ARTICLE{2017MNRAS.465.1454H,
       author = {{Hildebrandt}, H. and {Viola}, M. and {Heymans}, C. and {Joudaki}, S. and {Kuijken}, K. and {Blake}, C. and {Erben}, T. and {Joachimi}, B. and {Klaes}, D. and {Miller}, L. and {Morrison}, C.~B. and {Nakajima}, R. and {Verdoes Kleijn}, G. and {Amon}, A. and {Choi}, A. and {Covone}, G. and {de Jong}, J.~T.~A. and {Dvornik}, A. and {Fenech Conti}, I. and {Grado}, A. and {Harnois-D{\'e}raps}, J. and {Herbonnet}, R. and {Hoekstra}, H. and {K{\"o}hlinger}, F. and {McFarland}, J. and {Mead}, A. and {Merten}, J. and {Napolitano}, N. and {Peacock}, J.~A. and {Radovich}, M. and {Schneider}, P. and {Simon}, P. and {Valentijn}, E.~A. and {van den Busch}, J.~L. and {van Uitert}, E. and {Van Waerbeke}, L.},
        title = "{KiDS-450: cosmological parameter constraints from tomographic weak gravitational lensing}",
      journal = {\mnras},
         year = 2017,
        month = feb,
       volume = {465},
       number = {2},
        pages = {1454-1498},
          doi = {10.1093/mnras/stw2805},
archivePrefix = {arXiv},
       eprint = {1606.05338},
 primaryClass = {astro-ph.CO},
       adsurl = {https://ui.adsabs.harvard.edu/abs/2017MNRAS.465.1454H}
}

@ARTICLE{2021A&A...646A.140H,
       author = {{Heymans}, Catherine and {Tr{\"o}ster}, Tilman and {Asgari}, Marika and {Blake}, Chris and {Hildebrandt}, Hendrik and {Joachimi}, Benjamin and {Kuijken}, Konrad and {Lin}, Chieh-An and {S{\'a}nchez}, Ariel G. and {van den Busch}, Jan Luca and {Wright}, Angus H. and {Amon}, Alexandra and {Bilicki}, Maciej and {de Jong}, Jelte and {Crocce}, Martin and {Dvornik}, Andrej and {Erben}, Thomas and {Fortuna}, Maria Cristina and {Getman}, Fedor and {Giblin}, Benjamin and {Glazebrook}, Karl and {Hoekstra}, Henk and {Joudaki}, Shahab and {Kannawadi}, Arun and {K{\"o}hlinger}, Fabian and {Lidman}, Chris and {Miller}, Lance and {Napolitano}, Nicola R. and {Parkinson}, David and {Schneider}, Peter and {Shan}, HuanYuan and {Valentijn}, Edwin A. and {Verdoes Kleijn}, Gijs and {Wolf}, Christian},
        title = "{KiDS-1000 Cosmology: Multi-probe weak gravitational lensing and spectroscopic galaxy clustering constraints}",
      journal = {\aap},
         year = 2021,
        month = feb,
       volume = {646},
          eid = {A140},
        pages = {A140},
          doi = {10.1051/0004-6361/202039063},
archivePrefix = {arXiv},
       eprint = {2007.15632},
 primaryClass = {astro-ph.CO},
       adsurl = {https://ui.adsabs.harvard.edu/abs/2021A&A...646A.140H}
}

@ARTICLE{2018PhRvD..98d3526A,
       author = {{Abbott}, T.~M.~C. and {Abdalla}, F.~B. and {Alarcon}, A. and {Aleksi{\'c}}, J. and {Allam}, S. and {Allen}, S. and {Amara}, A. and {Annis}, J. and {Asorey}, J. and {Avila}, S. and {Bacon}, D. and {Balbinot}, E. and {Banerji}, M. and {Banik}, N. and {Barkhouse}, W. and {Baumer}, M. and {Baxter}, E. and {Bechtol}, K. and {Becker}, M.~R. and {Benoit-L{\'e}vy}, A. and {Benson}, B.~A. and {Bernstein}, G.~M. and {Bertin}, E. and {Blazek}, J. and {Bridle}, S.~L. and {Brooks}, D. and {Brout}, D. and {Buckley-Geer}, E. and {Burke}, D.~L. and {Busha}, M.~T. and {Campos}, A. and {Capozzi}, D. and {Carnero Rosell}, A. and {Carrasco Kind}, M. and {Carretero}, J. and {Castander}, F.~J. and {Cawthon}, R. and {Chang}, C. and {Chen}, N. and {Childress}, M. and {Choi}, A. and {Conselice}, C. and {Crittenden}, R. and {Crocce}, M. and {Cunha}, C.~E. and {D'Andrea}, C.~B. and {da Costa}, L.~N. and {Das}, R. and {Davis}, T.~M. and {Davis}, C. and {De Vicente}, J. and {DePoy}, D.~L. and {DeRose}, J. and {Desai}, S. and {Diehl}, H.~T. and {Dietrich}, J.~P. and {Dodelson}, S. and {Doel}, P. and {Drlica-Wagner}, A. and {Eifler}, T.~F. and {Elliott}, A.~E. and {Elsner}, F. and {Elvin-Poole}, J. and {Estrada}, J. and {Evrard}, A.~E. and {Fang}, Y. and {Fernandez}, E. and {Fert{\'e}}, A. and {Finley}, D.~A. and {Flaugher}, B. and {Fosalba}, P. and {Friedrich}, O. and {Frieman}, J. and {Garc{\'\i}a-Bellido}, J. and {Garcia-Fernandez}, M. and {Gatti}, M. and {Gaztanaga}, E. and {Gerdes}, D.~W. and {Giannantonio}, T. and {Gill}, M.~S.~S. and {Glazebrook}, K. and {Goldstein}, D.~A. and {Gruen}, D. and {Gruendl}, R.~A. and {Gschwend}, J. and {Gutierrez}, G. and {Hamilton}, S. and {Hartley}, W.~G. and {Hinton}, S.~R. and {Honscheid}, K. and {Hoyle}, B. and {Huterer}, D. and {Jain}, B. and {James}, D.~J. and {Jarvis}, M. and {Jeltema}, T. and {Johnson}, M.~D. and {Johnson}, M.~W.~G. and {Kacprzak}, T. and {Kent}, S. and {Kim}, A.~G. and {King}, A. and {Kirk}, D. and {Kokron}, N. and {Kovacs}, A. and {Krause}, E. and {Krawiec}, C. and {Kremin}, A. and {Kuehn}, K. and {Kuhlmann}, S. and {Kuropatkin}, N. and {Lacasa}, F. and {Lahav}, O. and {Li}, T.~S. and {Liddle}, A.~R. and {Lidman}, C. and {Lima}, M. and {Lin}, H. and {MacCrann}, N. and {Maia}, M.~A.~G. and {Makler}, M. and {Manera}, M. and {March}, M. and {Marshall}, J.~L. and {Martini}, P. and {McMahon}, R.~G. and {Melchior}, P. and {Menanteau}, F. and {Miquel}, R. and {Miranda}, V. and {Mudd}, D. and {Muir}, J. and {M{\"o}ller}, A. and {Neilsen}, E. and {Nichol}, R.~C. and {Nord}, B. and {Nugent}, P. and {Ogando}, R.~L.~C. and {Palmese}, A. and {Peacock}, J. and {Peiris}, H.~V. and {Peoples}, J. and {Percival}, W.~J. and {Petravick}, D. and {Plazas}, A.~A. and {Porredon}, A. and {Prat}, J. and {Pujol}, A. and {Rau}, M.~M. and {Refregier}, A. and {Ricker}, P.~M. and {Roe}, N. and {Rollins}, R.~P. and {Romer}, A.~K. and {Roodman}, A. and {Rosenfeld}, R. and {Ross}, A.~J. and {Rozo}, E. and {Rykoff}, E.~S. and {Sako}, M. and {Salvador}, A.~I. and {Samuroff}, S. and {S{\'a}nchez}, C. and {Sanchez}, E. and {Santiago}, B. and {Scarpine}, V. and {Schindler}, R. and {Scolnic}, D. and {Secco}, L.~F. and {Serrano}, S. and {Sevilla-Noarbe}, I. and {Sheldon}, E. and {Smith}, R.~C. and {Smith}, M. and {Smith}, J. and {Soares-Santos}, M. and {Sobreira}, F. and {Suchyta}, E. and {Tarle}, G. and {Thomas}, D. and {Troxel}, M.~A. and {Tucker}, D.~L. and {Tucker}, B.~E. and {Uddin}, S.~A. and {Varga}, T.~N. and {Vielzeuf}, P. and {Vikram}, V. and {Vivas}, A.~K. and {Walker}, A.~R. and {Wang}, M. and {Wechsler}, R.~H. and {Weller}, J. and {Wester}, W. and {Wolf}, R.~C. and {Yanny}, B. and {Yuan}, F. and {Zenteno}, A. and {Zhang}, B. and {Zhang}, Y. and {Zuntz}, J. and {Dark Energy Survey Collaboration}},
        title = "{Dark Energy Survey year 1 results: Cosmological constraints from galaxy clustering and weak lensing}",
      journal = {\prd},
         year = 2018,
        month = aug,
       volume = {98},
       number = {4},
          eid = {043526},
        pages = {043526},
          doi = {10.1103/PhysRevD.98.043526},
archivePrefix = {arXiv},
       eprint = {1708.01530},
 primaryClass = {astro-ph.CO},
       adsurl = {https://ui.adsabs.harvard.edu/abs/2018PhRvD..98d3526A}
}

@ARTICLE{2021arXiv210513549D,
       author = {{DES Collaboration} and {Abbott}, T.~M.~C. and {Aguena}, M. and {Alarcon}, A. and {Allam}, S. and {Alves}, O. and {Amon}, A. and {Andrade-Oliveira}, F. and {Annis}, J. and {Avila}, S. and {Bacon}, D. and {Baxter}, E. and {Bechtol}, K. and {Becker}, M.~R. and {Bernstein}, G.~M. and {Bhargava}, S. and {Birrer}, S. and {Blazek}, J. and {Brandao-Souza}, A. and {Bridle}, S.~L. and {Brooks}, D. and {Buckley-Geer}, E. and {Burke}, D.~L. and {Camacho}, H. and {Campos}, A. and {Carnero Rosell}, A. and {Carrasco Kind}, M. and {Carretero}, J. and {Castander}, F.~J. and {Cawthon}, R. and {Chang}, C. and {Chen}, A. and {Chen}, R. and {Choi}, A. and {Conselice}, C. and {Cordero}, J. and {Costanzi}, M. and {Crocce}, M. and {da Costa}, L.~N. and {da Silva Pereira}, M.~E. and {Davis}, C. and {Davis}, T.~M. and {De Vicente}, J. and {DeRose}, J. and {Desai}, S. and {Di Valentino}, E. and {Diehl}, H.~T. and {Dietrich}, J.~P. and {Dodelson}, S. and {Doel}, P. and {Doux}, C. and {Drlica-Wagner}, A. and {Eckert}, K. and {Eifler}, T.~F. and {Elsner}, F. and {Elvin-Poole}, J. and {Everett}, S. and {Evrard}, A.~E. and {Fang}, X. and {Farahi}, A. and {Fernandez}, E. and {Ferrero}, I. and {Fert{\'e}}, A. and {Fosalba}, P. and {Friedrich}, O. and {Frieman}, J. and {Garc{\'\i}a-Bellido}, J. and {Gatti}, M. and {Gaztanaga}, E. and {Gerdes}, D.~W. and {Giannantonio}, T. and {Giannini}, G. and {Gruen}, D. and {Gruendl}, R.~A. and {Gschwend}, J. and {Gutierrez}, G. and {Harrison}, I. and {Hartley}, W.~G. and {Herner}, K. and {Hinton}, S.~R. and {Hollowood}, D.~L. and {Honscheid}, K. and {Hoyle}, B. and {Huff}, E.~M. and {Huterer}, D. and {Jain}, B. and {James}, D.~J. and {Jarvis}, M. and {Jeffrey}, N. and {Jeltema}, T. and {Kovacs}, A. and {Krause}, E. and {Kron}, R. and {Kuehn}, K. and {Kuropatkin}, N. and {Lahav}, O. and {Leget}, P. -F. and {Lemos}, P. and {Liddle}, A.~R. and {Lidman}, C. and {Lima}, M. and {Lin}, H. and {MacCrann}, N. and {Maia}, M.~A.~G. and {Marshall}, J.~L. and {Martini}, P. and {McCullough}, J. and {Melchior}, P. and {Mena-Fern{\'a}ndez}, J. and {Menanteau}, F. and {Miquel}, R. and {Mohr}, J.~J. and {Morgan}, R. and {Muir}, J. and {Myles}, J. and {Nadathur}, S. and {Navarro-Alsina}, A. and {Nichol}, R.~C. and {Ogando}, R.~L.~C. and {Omori}, Y. and {Palmese}, A. and {Pandey}, S. and {Park}, Y. and {Paz-Chinch{\'o}n}, F. and {Petravick}, D. and {Pieres}, A. and {Plazas Malag{\'o}n}, A.~A. and {Porredon}, A. and {Prat}, J. and {Raveri}, M. and {Rodriguez-Monroy}, M. and {Rollins}, R.~P. and {Romer}, A.~K. and {Roodman}, A. and {Rosenfeld}, R. and {Ross}, A.~J. and {Rykoff}, E.~S. and {Samuroff}, S. and {S{\'a}nchez}, C. and {Sanchez}, E. and {Sanchez}, J. and {Sanchez Cid}, D. and {Scarpine}, V. and {Schubnell}, M. and {Scolnic}, D. and {Secco}, L.~F. and {Serrano}, S. and {Sevilla-Noarbe}, I. and {Sheldon}, E. and {Shin}, T. and {Smith}, M. and {Soares-Santos}, M. and {Suchyta}, E. and {Swanson}, M.~E.~C. and {Tabbutt}, M. and {Tarle}, G. and {Thomas}, D. and {To}, C. and {Troja}, A. and {Troxel}, M.~A. and {Tucker}, D.~L. and {Tutusaus}, I. and {Varga}, T.~N. and {Walker}, A.~R. and {Weaverdyck}, N. and {Weller}, J. and {Yanny}, B. and {Yin}, B. and {Zhang}, Y. and {Zuntz}, J.},
        title = "{Dark Energy Survey Year 3 Results: Cosmological Constraints from Galaxy Clustering and Weak Lensing}",
      journal = {arXiv e-prints},
         year = 2021,
        month = may,
          eid = {arXiv:2105.13549},
        pages = {arXiv:2105.13549},
archivePrefix = {arXiv},
       eprint = {2105.13549},
 primaryClass = {astro-ph.CO},
       adsurl = {https://ui.adsabs.harvard.edu/abs/2021arXiv210513549D}
}

@ARTICLE{2019PASJ...71...43H,
       author = {{Hikage}, Chiaki and {Oguri}, Masamune and {Hamana}, Takashi and {More}, Surhud and {Mandelbaum}, Rachel and {Takada}, Masahiro and {K{\"o}hlinger}, Fabian and {Miyatake}, Hironao and {Nishizawa}, Atsushi J. and {Aihara}, Hiroaki and {Armstrong}, Robert and {Bosch}, James and {Coupon}, Jean and {Ducout}, Anne and {Ho}, Paul and {Hsieh}, Bau-Ching and {Komiyama}, Yutaka and {Lanusse}, Fran{\c{c}}ois and {Leauthaud}, Alexie and {Lupton}, Robert H. and {Medezinski}, Elinor and {Mineo}, Sogo and {Miyama}, Shoken and {Miyazaki}, Satoshi and {Murata}, Ryoma and {Murayama}, Hitoshi and {Shirasaki}, Masato and {Sif{\'o}n}, Crist{\'o}bal and {Simet}, Melanie and {Speagle}, Joshua and {Spergel}, David N. and {Strauss}, Michael A. and {Sugiyama}, Naoshi and {Tanaka}, Masayuki and {Utsumi}, Yousuke and {Wang}, Shiang-Yu and {Yamada}, Yoshihiko},
        title = "{Cosmology from cosmic shear power spectra with Subaru Hyper Suprime-Cam first-year data}",
      journal = {\pasj},
         year = 2019,
        month = apr,
       volume = {71},
       number = {2},
          eid = {43},
        pages = {43},
          doi = {10.1093/pasj/psz010},
archivePrefix = {arXiv},
       eprint = {1809.09148},
 primaryClass = {astro-ph.CO},
       adsurl = {https://ui.adsabs.harvard.edu/abs/2019PASJ...71...43H}
}

@ARTICLE{2016arXiv161100036D,
       author = {{DESI Collaboration} and {Aghamousa}, Amir and {Aguilar}, Jessica and {Ahlen}, Steve and {Alam}, Shadab and {Allen}, Lori E. and {Allende Prieto}, Carlos and {Annis}, James and {Bailey}, Stephen and {Balland}, Christophe and {Ballester}, Otger and {Baltay}, Charles and {Beaufore}, Lucas and {Bebek}, Chris and {Beers}, Timothy C. and {Bell}, Eric F. and {Bernal}, Jos{\'e} Luis and {Besuner}, Robert and {Beutler}, Florian and {Blake}, Chris and {Bleuler}, Hannes and {Blomqvist}, Michael and {Blum}, Robert and {Bolton}, Adam S. and {Briceno}, Cesar and {Brooks}, David and {Brownstein}, Joel R. and {Buckley-Geer}, Elizabeth and {Burden}, Angela and {Burtin}, Etienne and {Busca}, Nicolas G. and {Cahn}, Robert N. and {Cai}, Yan-Chuan and {Cardiel-Sas}, Laia and {Carlberg}, Raymond G. and {Carton}, Pierre-Henri and {Casas}, Ricard and {Castander}, Francisco J. and {Cervantes-Cota}, Jorge L. and {Claybaugh}, Todd M. and {Close}, Madeline and {Coker}, Carl T. and {Cole}, Shaun and {Comparat}, Johan and {Cooper}, Andrew P. and {Cousinou}, M. -C. and {Crocce}, Martin and {Cuby}, Jean-Gabriel and {Cunningham}, Daniel P. and {Davis}, Tamara M. and {Dawson}, Kyle S. and {de la Macorra}, Axel and {De Vicente}, Juan and {Delubac}, Timoth{\'e}e and {Derwent}, Mark and {Dey}, Arjun and {Dhungana}, Govinda and {Ding}, Zhejie and {Doel}, Peter and {Duan}, Yutong T. and {Ealet}, Anne and {Edelstein}, Jerry and {Eftekharzadeh}, Sarah and {Eisenstein}, Daniel J. and {Elliott}, Ann and {Escoffier}, St{\'e}phanie and {Evatt}, Matthew and {Fagrelius}, Parker and {Fan}, Xiaohui and {Fanning}, Kevin and {Farahi}, Arya and {Farihi}, Jay and {Favole}, Ginevra and {Feng}, Yu and {Fernandez}, Enrique and {Findlay}, Joseph R. and {Finkbeiner}, Douglas P. and {Fitzpatrick}, Michael J. and {Flaugher}, Brenna and {Flender}, Samuel and {Font-Ribera}, Andreu and {Forero-Romero}, Jaime E. and {Fosalba}, Pablo and {Frenk}, Carlos S. and {Fumagalli}, Michele and {Gaensicke}, Boris T. and {Gallo}, Giuseppe and {Garcia-Bellido}, Juan and {Gaztanaga}, Enrique and {Pietro Gentile Fusillo}, Nicola and {Gerard}, Terry and {Gershkovich}, Irena and {Giannantonio}, Tommaso and {Gillet}, Denis and {Gonzalez-de-Rivera}, Guillermo and {Gonzalez-Perez}, Violeta and {Gott}, Shelby and {Graur}, Or and {Gutierrez}, Gaston and {Guy}, Julien and {Habib}, Salman and {Heetderks}, Henry and {Heetderks}, Ian and {Heitmann}, Katrin and {Hellwing}, Wojciech A. and {Herrera}, David A. and {Ho}, Shirley and {Holland}, Stephen and {Honscheid}, Klaus and {Huff}, Eric and {Hutchinson}, Timothy A. and {Huterer}, Dragan and {Hwang}, Ho Seong and {Illa Laguna}, Joseph Maria and {Ishikawa}, Yuzo and {Jacobs}, Dianna and {Jeffrey}, Niall and {Jelinsky}, Patrick and {Jennings}, Elise and {Jiang}, Linhua and {Jimenez}, Jorge and {Johnson}, Jennifer and {Joyce}, Richard and {Jullo}, Eric and {Juneau}, St{\'e}phanie and {Kama}, Sami and {Karcher}, Armin and {Karkar}, Sonia and {Kehoe}, Robert and {Kennamer}, Noble and {Kent}, Stephen and {Kilbinger}, Martin and {Kim}, Alex G. and {Kirkby}, David and {Kisner}, Theodore and {Kitanidis}, Ellie and {Kneib}, Jean-Paul and {Koposov}, Sergey and {Kovacs}, Eve and {Koyama}, Kazuya and {Kremin}, Anthony and {Kron}, Richard and {Kronig}, Luzius and {Kueter-Young}, Andrea and {Lacey}, Cedric G. and {Lafever}, Robin and {Lahav}, Ofer and {Lambert}, Andrew and {Lampton}, Michael and {Landriau}, Martin and {Lang}, Dustin and {Lauer}, Tod R. and {Le Goff}, Jean-Marc and {Le Guillou}, Laurent and {Le Van Suu}, Auguste and {Lee}, Jae Hyeon and {Lee}, Su-Jeong and {Leitner}, Daniela and {Lesser}, Michael and {Levi}, Michael E. and {L'Huillier}, Benjamin and {Li}, Baojiu and {Liang}, Ming and {Lin}, Huan and {Linder}, Eric and {Loebman}, Sarah R. and {Luki{\'c}}, Zarija and {Ma}, Jun and {MacCrann}, Niall and {Magneville}, Christophe and {Makarem}, Laleh and {Manera}, Marc and {Manser}, Christopher J. and {Marshall}, Robert and {Martini}, Paul and {Massey}, Richard and {Matheson}, Thomas and {McCauley}, Jeremy and {McDonald}, Patrick and {McGreer}, Ian D. and {Meisner}, Aaron and {Metcalfe}, Nigel and {Miller}, Timothy N. and {Miquel}, Ramon and {Moustakas}, John and {Myers}, Adam and {Naik}, Milind and {Newman}, Jeffrey A. and {Nichol}, Robert C. and {Nicola}, Andrina and {Nicolati da Costa}, Luiz and {Nie}, Jundan and {Niz}, Gustavo and {Norberg}, Peder and {Nord}, Brian and {Norman}, Dara and {Nugent}, Peter and {O'Brien}, Thomas and {Oh}, Minji and {Olsen}, Knut A.~G. and {Padilla}, Cristobal and {Padmanabhan}, Hamsa and {Padmanabhan}, Nikhil and {Palanque-Delabrouille}, Nathalie and {Palmese}, Antonella and {Pappalardo}, Daniel and {P{\^a}ris}, Isabelle and {Park}, Changbom and {Patej}, Anna and {Peacock}, John A. and {Peiris}, Hiranya V. and {Peng}, Xiyan and {Percival}, Will J. and {Perruchot}, Sandrine and {Pieri}, Matthew M. and {Pogge}, Richard and {Pollack}, Jennifer E. and {Poppett}, Claire and {Prada}, Francisco and {Prakash}, Abhishek and {Probst}, Ronald G. and {Rabinowitz}, David and {Raichoor}, Anand and {Ree}, Chang Hee and {Refregier}, Alexandre and {Regal}, Xavier and {Reid}, Beth and {Reil}, Kevin and {Rezaie}, Mehdi and {Rockosi}, Constance M. and {Roe}, Natalie and {Ronayette}, Samuel and {Roodman}, Aaron and {Ross}, Ashley J. and {Ross}, Nicholas P. and {Rossi}, Graziano and {Rozo}, Eduardo and {Ruhlmann-Kleider}, Vanina and {Rykoff}, Eli S. and {Sabiu}, Cristiano and {Samushia}, Lado and {Sanchez}, Eusebio and {Sanchez}, Javier and {Schlegel}, David J. and {Schneider}, Michael and {Schubnell}, Michael and {Secroun}, Aur{\'e}lia and {Seljak}, Uros and {Seo}, Hee-Jong and {Serrano}, Santiago and {Shafieloo}, Arman and {Shan}, Huanyuan and {Sharples}, Ray and {Sholl}, Michael J. and {Shourt}, William V. and {Silber}, Joseph H. and {Silva}, David R. and {Sirk}, Martin M. and {Slosar}, Anze and {Smith}, Alex and {Smoot}, George F. and {Som}, Debopam and {Song}, Yong-Seon and {Sprayberry}, David and {Staten}, Ryan and {Stefanik}, Andy and {Tarle}, Gregory and {Sien Tie}, Suk and {Tinker}, Jeremy L. and {Tojeiro}, Rita and {Valdes}, Francisco and {Valenzuela}, Octavio and {Valluri}, Monica and {Vargas-Magana}, Mariana and {Verde}, Licia and {Walker}, Alistair R. and {Wang}, Jiali and {Wang}, Yuting and {Weaver}, Benjamin A. and {Weaverdyck}, Curtis and {Wechsler}, Risa H. and {Weinberg}, David H. and {White}, Martin and {Yang}, Qian and {Yeche}, Christophe and {Zhang}, Tianmeng and {Zhao}, Gong-Bo and {Zheng}, Yi and {Zhou}, Xu and {Zhou}, Zhimin and {Zhu}, Yaling and {Zou}, Hu and {Zu}, Ying},
        title = "{The DESI Experiment Part I: Science,Targeting, and Survey Design}",
      journal = {arXiv e-prints},
         year = 2016,
        month = oct,
          eid = {arXiv:1611.00036},
        pages = {arXiv:1611.00036},
archivePrefix = {arXiv},
       eprint = {1611.00036},
 primaryClass = {astro-ph.IM},
       adsurl = {https://ui.adsabs.harvard.edu/abs/2016arXiv161100036D}
}

@ARTICLE{2019ApJ...873..111I,
       author = {{Ivezi{\'c}}, {\v{Z}}eljko and {Kahn}, Steven M. and {Tyson}, J. Anthony and {Abel}, Bob and {Acosta}, Emily and {Allsman}, Robyn and {Alonso}, David and {AlSayyad}, Yusra and {Anderson}, Scott F. and {Andrew}, John and {Angel}, James Roger P. and {Angeli}, George Z. and {Ansari}, Reza and {Antilogus}, Pierre and {Araujo}, Constanza and {Armstrong}, Robert and {Arndt}, Kirk T. and {Astier}, Pierre and {Aubourg}, {\'E}ric and {Auza}, Nicole and {Axelrod}, Tim S. and {Bard}, Deborah J. and {Barr}, Jeff D. and {Barrau}, Aurelian and {Bartlett}, James G. and {Bauer}, Amanda E. and {Bauman}, Brian J. and {Baumont}, Sylvain and {Bechtol}, Ellen and {Bechtol}, Keith and {Becker}, Andrew C. and {Becla}, Jacek and {Beldica}, Cristina and {Bellavia}, Steve and {Bianco}, Federica B. and {Biswas}, Rahul and {Blanc}, Guillaume and {Blazek}, Jonathan and {Blandford}, Roger D. and {Bloom}, Josh S. and {Bogart}, Joanne and {Bond}, Tim W. and {Booth}, Michael T. and {Borgland}, Anders W. and {Borne}, Kirk and {Bosch}, James F. and {Boutigny}, Dominique and {Brackett}, Craig A. and {Bradshaw}, Andrew and {Brandt}, William Nielsen and {Brown}, Michael E. and {Bullock}, James S. and {Burchat}, Patricia and {Burke}, David L. and {Cagnoli}, Gianpietro and {Calabrese}, Daniel and {Callahan}, Shawn and {Callen}, Alice L. and {Carlin}, Jeffrey L. and {Carlson}, Erin L. and {Chandrasekharan}, Srinivasan and {Charles-Emerson}, Glenaver and {Chesley}, Steve and {Cheu}, Elliott C. and {Chiang}, Hsin-Fang and {Chiang}, James and {Chirino}, Carol and {Chow}, Derek and {Ciardi}, David R. and {Claver}, Charles F. and {Cohen-Tanugi}, Johann and {Cockrum}, Joseph J. and {Coles}, Rebecca and {Connolly}, Andrew J. and {Cook}, Kem H. and {Cooray}, Asantha and {Covey}, Kevin R. and {Cribbs}, Chris and {Cui}, Wei and {Cutri}, Roc and {Daly}, Philip N. and {Daniel}, Scott F. and {Daruich}, Felipe and {Daubard}, Guillaume and {Daues}, Greg and {Dawson}, William and {Delgado}, Francisco and {Dellapenna}, Alfred and {de Peyster}, Robert and {de Val-Borro}, Miguel and {Digel}, Seth W. and {Doherty}, Peter and {Dubois}, Richard and {Dubois-Felsmann}, Gregory P. and {Durech}, Josef and {Economou}, Frossie and {Eifler}, Tim and {Eracleous}, Michael and {Emmons}, Benjamin L. and {Fausti Neto}, Angelo and {Ferguson}, Henry and {Figueroa}, Enrique and {Fisher-Levine}, Merlin and {Focke}, Warren and {Foss}, Michael D. and {Frank}, James and {Freemon}, Michael D. and {Gangler}, Emmanuel and {Gawiser}, Eric and {Geary}, John C. and {Gee}, Perry and {Geha}, Marla and {Gessner}, Charles J.~B. and {Gibson}, Robert R. and {Gilmore}, D. Kirk and {Glanzman}, Thomas and {Glick}, William and {Goldina}, Tatiana and {Goldstein}, Daniel A. and {Goodenow}, Iain and {Graham}, Melissa L. and {Gressler}, William J. and {Gris}, Philippe and {Guy}, Leanne P. and {Guyonnet}, Augustin and {Haller}, Gunther and {Harris}, Ron and {Hascall}, Patrick A. and {Haupt}, Justine and {Hernandez}, Fabio and {Herrmann}, Sven and {Hileman}, Edward and {Hoblitt}, Joshua and {Hodgson}, John A. and {Hogan}, Craig and {Howard}, James D. and {Huang}, Dajun and {Huffer}, Michael E. and {Ingraham}, Patrick and {Innes}, Walter R. and {Jacoby}, Suzanne H. and {Jain}, Bhuvnesh and {Jammes}, Fabrice and {Jee}, M. James and {Jenness}, Tim and {Jernigan}, Garrett and {Jevremovi{\'c}}, Darko and {Johns}, Kenneth and {Johnson}, Anthony S. and {Johnson}, Margaret W.~G. and {Jones}, R. Lynne and {Juramy-Gilles}, Claire and {Juri{\'c}}, Mario and {Kalirai}, Jason S. and {Kallivayalil}, Nitya J. and {Kalmbach}, Bryce and {Kantor}, Jeffrey P. and {Karst}, Pierre and {Kasliwal}, Mansi M. and {Kelly}, Heather and {Kessler}, Richard and {Kinnison}, Veronica and {Kirkby}, David and {Knox}, Lloyd and {Kotov}, Ivan V. and {Krabbendam}, Victor L. and {Krughoff}, K. Simon and {Kub{\'a}nek}, Petr and {Kuczewski}, John and {Kulkarni}, Shri and {Ku}, John and {Kurita}, Nadine R. and {Lage}, Craig S. and {Lambert}, Ron and {Lange}, Travis and {Langton}, J. Brian and {Le Guillou}, Laurent and {Levine}, Deborah and {Liang}, Ming and {Lim}, Kian-Tat and {Lintott}, Chris J. and {Long}, Kevin E. and {Lopez}, Margaux and {Lotz}, Paul J. and {Lupton}, Robert H. and {Lust}, Nate B. and {MacArthur}, Lauren A. and {Mahabal}, Ashish and {Mandelbaum}, Rachel and {Markiewicz}, Thomas W. and {Marsh}, Darren S. and {Marshall}, Philip J. and {Marshall}, Stuart and {May}, Morgan and {McKercher}, Robert and {McQueen}, Michelle and {Meyers}, Joshua and {Migliore}, Myriam and {Miller}, Michelle and {Mills}, David J. and {Miraval}, Connor and {Moeyens}, Joachim and {Moolekamp}, Fred E. and {Monet}, David G. and {Moniez}, Marc and {Monkewitz}, Serge and {Montgomery}, Christopher and {Morrison}, Christopher B. and {Mueller}, Fritz and {Muller}, Gary P. and {Mu{\~n}oz Arancibia}, Freddy and {Neill}, Douglas R. and {Newbry}, Scott P. and {Nief}, Jean-Yves and {Nomerotski}, Andrei and {Nordby}, Martin and {O'Connor}, Paul and {Oliver}, John and {Olivier}, Scot S. and {Olsen}, Knut and {O'Mullane}, William and {Ortiz}, Sandra and {Osier}, Shawn and {Owen}, Russell E. and {Pain}, Reynald and {Palecek}, Paul E. and {Parejko}, John K. and {Parsons}, James B. and {Pease}, Nathan M. and {Peterson}, J. Matt and {Peterson}, John R. and {Petravick}, Donald L. and {Libby Petrick}, M.~E. and {Petry}, Cathy E. and {Pierfederici}, Francesco and {Pietrowicz}, Stephen and {Pike}, Rob and {Pinto}, Philip A. and {Plante}, Raymond and {Plate}, Stephen and {Plutchak}, Joel P. and {Price}, Paul A. and {Prouza}, Michael and {Radeka}, Veljko and {Rajagopal}, Jayadev and {Rasmussen}, Andrew P. and {Regnault}, Nicolas and {Reil}, Kevin A. and {Reiss}, David J. and {Reuter}, Michael A. and {Ridgway}, Stephen T. and {Riot}, Vincent J. and {Ritz}, Steve and {Robinson}, Sean and {Roby}, William and {Roodman}, Aaron and {Rosing}, Wayne and {Roucelle}, Cecille and {Rumore}, Matthew R. and {Russo}, Stefano and {Saha}, Abhijit and {Sassolas}, Benoit and {Schalk}, Terry L. and {Schellart}, Pim and {Schindler}, Rafe H. and {Schmidt}, Samuel and {Schneider}, Donald P. and {Schneider}, Michael D. and {Schoening}, William and {Schumacher}, German and {Schwamb}, Megan E. and {Sebag}, Jacques and {Selvy}, Brian and {Sembroski}, Glenn H. and {Seppala}, Lynn G. and {Serio}, Andrew and {Serrano}, Eduardo and {Shaw}, Richard A. and {Shipsey}, Ian and {Sick}, Jonathan and {Silvestri}, Nicole and {Slater}, Colin T. and {Smith}, J. Allyn and {Smith}, R. Chris and {Sobhani}, Shahram and {Soldahl}, Christine and {Storrie-Lombardi}, Lisa and {Stover}, Edward and {Strauss}, Michael A. and {Street}, Rachel A. and {Stubbs}, Christopher W. and {Sullivan}, Ian S. and {Sweeney}, Donald and {Swinbank}, John D. and {Szalay}, Alexander and {Takacs}, Peter and {Tether}, Stephen A. and {Thaler}, Jon J. and {Thayer}, John Gregg and {Thomas}, Sandrine and {Thornton}, Adam J. and {Thukral}, Vaikunth and {Tice}, Jeffrey and {Trilling}, David E. and {Turri}, Max and {Van Berg}, Richard and {Vanden Berk}, Daniel and {Vetter}, Kurt and {Virieux}, Francoise and {Vucina}, Tomislav and {Wahl}, William and {Walkowicz}, Lucianne and {Walsh}, Brian and {Walter}, Christopher W. and {Wang}, Daniel L. and {Wang}, Shin-Yawn and {Warner}, Michael and {Wiecha}, Oliver and {Willman}, Beth and {Winters}, Scott E. and {Wittman}, David and {Wolff}, Sidney C. and {Wood-Vasey}, W. Michael and {Wu}, Xiuqin and {Xin}, Bo and {Yoachim}, Peter and {Zhan}, Hu},
        title = "{LSST: From Science Drivers to Reference Design and Anticipated Data Products}",
      journal = {\apj},
         year = 2019,
        month = mar,
       volume = {873},
       number = {2},
          eid = {111},
        pages = {111},
          doi = {10.3847/1538-4357/ab042c},
archivePrefix = {arXiv},
       eprint = {0805.2366},
 primaryClass = {astro-ph},
       adsurl = {https://ui.adsabs.harvard.edu/abs/2019ApJ...873..111I}
}

@ARTICLE{2019arXiv190205569A,
       author = {{Akeson}, Rachel and {Armus}, Lee and {Bachelet}, Etienne and {Bailey}, Vanessa and {Bartusek}, Lisa and {Bellini}, Andrea and {Benford}, Dominic and {Bennett}, David and {Bhattacharya}, Aparna and {Bohlin}, Ralph and {Boyer}, Martha and {Bozza}, Valerio and {Bryden}, Geoffrey and {Calchi Novati}, Sebastiano and {Carpenter}, Kenneth and {Casertano}, Stefano and {Choi}, Ami and {Content}, David and {Dayal}, Pratika and {Dressler}, Alan and {Dor{\'e}}, Olivier and {Fall}, S. Michael and {Fan}, Xiaohui and {Fang}, Xiao and {Filippenko}, Alexei and {Finkelstein}, Steven and {Foley}, Ryan and {Furlanetto}, Steven and {Kalirai}, Jason and {Gaudi}, B. Scott and {Gilbert}, Karoline and {Girard}, Julien and {Grady}, Kevin and {Greene}, Jenny and {Guhathakurta}, Puragra and {Heinrich}, Chen and {Hemmati}, Shoubaneh and {Hendel}, David and {Henderson}, Calen and {Henning}, Thomas and {Hirata}, Christopher and {Ho}, Shirley and {Huff}, Eric and {Hutter}, Anne and {Jansen}, Rolf and {Jha}, Saurabh and {Johnson}, Samson and {Jones}, David and {Kasdin}, Jeremy and {Kelly}, Patrick and {Kirshner}, Robert and {Koekemoer}, Anton and {Kruk}, Jeffrey and {Lewis}, Nikole and {Macintosh}, Bruce and {Madau}, Piero and {Malhotra}, Sangeeta and {Mandel}, Kaisey and {Massara}, Elena and {Masters}, Daniel and {McEnery}, Julie and {McQuinn}, Kristen and {Melchior}, Peter and {Melton}, Mark and {Mennesson}, Bertrand and {Peeples}, Molly and {Penny}, Matthew and {Perlmutter}, Saul and {Pisani}, Alice and {Plazas}, Andr{\'e}s and {Poleski}, Radek and {Postman}, Marc and {Ranc}, Cl{\'e}ment and {Rauscher}, Bernard and {Rest}, Armin and {Roberge}, Aki and {Robertson}, Brant and {Rodney}, Steven and {Rhoads}, James and {Rhodes}, Jason and {Ryan}, Russell, Jr. and {Sahu}, Kailash and {Sand}, David and {Scolnic}, Dan and {Seth}, Anil and {Shvartzvald}, Yossi and {Siellez}, Karelle and {Smith}, Arfon and {Spergel}, David and {Stassun}, Keivan and {Street}, Rachel and {Strolger}, Louis-Gregory and {Szalay}, Alexander and {Trauger}, John and {Troxel}, M.~A. and {Turnbull}, Margaret and {van der Marel}, Roeland and {von der Linden}, Anja and {Wang}, Yun and {Weinberg}, David and {Williams}, Benjamin and {Windhorst}, Rogier and {Wollack}, Edward and {Wu}, Hao-Yi and {Yee}, Jennifer and {Zimmerman}, Neil},
        title = "{The Wide Field Infrared Survey Telescope: 100 Hubbles for the 2020s}",
      journal = {arXiv e-prints},
         year = 2019,
        month = feb,
          eid = {arXiv:1902.05569},
        pages = {arXiv:1902.05569},
archivePrefix = {arXiv},
       eprint = {1902.05569},
 primaryClass = {astro-ph.IM},
       adsurl = {https://ui.adsabs.harvard.edu/abs/2019arXiv190205569A}
}

@ARTICLE{Euclid_WhitePaper,
       author = {{Laureijs}, R. and {Amiaux}, J. and {Arduini}, S. and
         {Augu{\`e}res}, J. -L. and {Brinchmann}, J. and {Cole}, R. and
         {Cropper}, M. and {Dabin}, C. and {Duvet}, L. and {Ealet}, A. and
         {Garilli}, B. and {Gondoin}, P. and {Guzzo}, L. and {Hoar}, J. and
         {Hoekstra}, H. and {Holmes}, R. and {Kitching}, T. and {Maciaszek}, T. and
         {Mellier}, Y. and {Pasian}, F. and {Percival}, W. and {Rhodes}, J. and
         {Saavedra Criado}, G. and {Sauvage}, M. and {Scaramella}, R. and
         {Valenziano}, L. and {Warren}, S. and {Bender}, R. and {Castander}, F. and
         {Cimatti}, A. and {Le F{\`e}vre}, O. and {Kurki-Suonio}, H. and
         {Levi}, M. and {Lilje}, P. and {Meylan}, G. and {Nichol}, R. and
         {Pedersen}, K. and {Popa}, V. and {Rebolo Lopez}, R. and {Rix}, H. -W. and
         {Rottgering}, H. and {Zeilinger}, W. and {Grupp}, F. and {Hudelot}, P. and
         {Massey}, R. and {Meneghetti}, M. and {Miller}, L. and {Paltani}, S. and
         {Paulin-Henriksson}, S. and {Pires}, S. and {Saxton}, C. and
         {Schrabback}, T. and {Seidel}, G. and {Walsh}, J. and {Aghanim}, N. and
         {Amendola}, L. and {Bartlett}, J. and {Baccigalupi}, C. and
         {Beaulieu}, J. -P. and {Benabed}, K. and {Cuby}, J. -G. and
         {Elbaz}, D. and {Fosalba}, P. and {Gavazzi}, G. and {Helmi}, A. and
         {Hook}, I. and {Irwin}, M. and {Kneib}, J. -P. and {Kunz}, M. and
         {Mannucci}, F. and {Moscardini}, L. and {Tao}, C. and {Teyssier}, R. and
         {Weller}, J. and {Zamorani}, G. and {Zapatero Osorio}, M.~R. and
         {Boulade}, O. and {Foumond}, J.~J. and {Di Giorgio}, A. and
         {Guttridge}, P. and {James}, A. and {Kemp}, M. and {Martignac}, J. and
         {Spencer}, A. and {Walton}, D. and {Bl{\"u}mchen}, T. and {Bonoli}, C. and
         {Bortoletto}, F. and {Cerna}, C. and {Corcione}, L. and {Fabron}, C. and
         {Jahnke}, K. and {Ligori}, S. and {Madrid}, F. and {Martin}, L. and
         {Morgante}, G. and {Pamplona}, T. and {Prieto}, E. and {Riva}, M. and
         {Toledo}, R. and {Trifoglio}, M. and {Zerbi}, F. and {Abdalla}, F. and
         {Douspis}, M. and {Grenet}, C. and {Borgani}, S. and {Bouwens}, R. and
         {Courbin}, F. and {Delouis}, J. -M. and {Dubath}, P. and {Fontana}, A. and
         {Frailis}, M. and {Grazian}, A. and {Koppenh{\"o}fer}, J. and
         {Mansutti}, O. and {Melchior}, M. and {Mignoli}, M. and {Mohr}, J. and
         {Neissner}, C. and {Noddle}, K. and {Poncet}, M. and {Scodeggio}, M. and
         {Serrano}, S. and {Shane}, N. and {Starck}, J. -L. and {Surace}, C. and
         {Taylor}, A. and {Verdoes-Kleijn}, G. and {Vuerli}, C. and
         {Williams}, O.~R. and {Zacchei}, A. and {Altieri}, B. and
         {Escudero Sanz}, I. and {Kohley}, R. and {Oosterbroek}, T. and
         {Astier}, P. and {Bacon}, D. and {Bardelli}, S. and {Baugh}, C. and
         {Bellagamba}, F. and {Benoist}, C. and {Bianchi}, D. and {Biviano}, A. and
         {Branchini}, E. and {Carbone}, C. and {Cardone}, V. and {Clements}, D. and
         {Colombi}, S. and {Conselice}, C. and {Cresci}, G. and {Deacon}, N. and
         {Dunlop}, J. and {Fedeli}, C. and {Fontanot}, F. and {Franzetti}, P. and
         {Giocoli}, C. and {Garcia-Bellido}, J. and {Gow}, J. and {Heavens}, A. and
         {Hewett}, P. and {Heymans}, C. and {Holland}, A. and {Huang}, Z. and
         {Ilbert}, O. and {Joachimi}, B. and {Jennins}, E. and {Kerins}, E. and
         {Kiessling}, A. and {Kirk}, D. and {Kotak}, R. and {Krause}, O. and
         {Lahav}, O. and {van Leeuwen}, F. and {Lesgourgues}, J. and
         {Lombardi}, M. and {Magliocchetti}, M. and {Maguire}, K. and
         {Majerotto}, E. and {Maoli}, R. and {Marulli}, F. and
         {Maurogordato}, S. and {McCracken}, H. and {McLure}, R. and
         {Melchiorri}, A. and {Merson}, A. and {Moresco}, M. and {Nonino}, M. and
         {Norberg}, P. and {Peacock}, J. and {Pello}, R. and {Penny}, M. and
         {Pettorino}, V. and {Di Porto}, C. and {Pozzetti}, L. and
         {Quercellini}, C. and {Radovich}, M. and {Rassat}, A. and {Roche}, N. and
         {Ronayette}, S. and {Rossetti}, E. and {Sartoris}, B. and
         {Schneider}, P. and {Semboloni}, E. and {Serjeant}, S. and
         {Simpson}, F. and {Skordis}, C. and {Smadja}, G. and {Smartt}, S. and
         {Spano}, P. and {Spiro}, S. and {Sullivan}, M. and {Tilquin}, A. and
         {Trotta}, R. and {Verde}, L. and {Wang}, Y. and {Williger}, G. and
         {Zhao}, G. and {Zoubian}, J. and {Zucca}, E.},
        title = "{Euclid Definition Study Report}",
      journal = {arXiv e-prints},
         year = "2011",
        month = "Oct",
          eid = {arXiv:1110.3193},
        pages = {arXiv:1110.3193},
archivePrefix = {arXiv},
       eprint = {1110.3193},
 primaryClass = {astro-ph.CO},
       adsurl = {https://ui.adsabs.harvard.edu/abs/2011arXiv1110.3193L}
}

@ARTICLE{2018PhR...733....1D,
       author = {{Desjacques}, Vincent and {Jeong}, Donghui and {Schmidt}, Fabian},
        title = "{Large-scale galaxy bias}",
      journal = {\physrep},
         year = 2018,
        month = feb,
       volume = {733},
        pages = {1-193},
          doi = {10.1016/j.physrep.2017.12.002},
archivePrefix = {arXiv},
       eprint = {1611.09787},
 primaryClass = {astro-ph.CO},
       adsurl = {https://ui.adsabs.harvard.edu/abs/2018PhR...733....1D}
}

@ARTICLE{2014arXiv1412.4872D,
       author = {{Dor{\'e}}, Olivier and {Bock}, Jamie and {Ashby}, Matthew and {Capak}, Peter and {Cooray}, Asantha and {de Putter}, Roland and {Eifler}, Tim and {Flagey}, Nicolas and {Gong}, Yan and {Habib}, Salman and {Heitmann}, Katrin and {Hirata}, Chris and {Jeong}, Woong-Seob and {Katti}, Raj and {Korngut}, Phil and {Krause}, Elisabeth and {Lee}, Dae-Hee and {Masters}, Daniel and {Mauskopf}, Phil and {Melnick}, Gary and {Mennesson}, Bertrand and {Nguyen}, Hien and {{\"O}berg}, Karin and {Pullen}, Anthony and {Raccanelli}, Alvise and {Smith}, Roger and {Song}, Yong-Seon and {Tolls}, Volker and {Unwin}, Steve and {Venumadhav}, Tejaswi and {Viero}, Marco and {Werner}, Mike and {Zemcov}, Mike},
        title = "{Cosmology with the SPHEREX All-Sky Spectral Survey}",
      journal = {arXiv e-prints},
         year = 2014,
        month = dec,
          eid = {arXiv:1412.4872},
        pages = {arXiv:1412.4872},
archivePrefix = {arXiv},
       eprint = {1412.4872},
 primaryClass = {astro-ph.CO},
       adsurl = {https://ui.adsabs.harvard.edu/abs/2014arXiv1412.4872D}
}

@article{kaiser:84,
  	author = {{Kaiser}, N.},
    	title = {{On the spatial correlations of Abell clusters}},
  	journal = {\apjl},
     	year = 1984,
    	month = sep,
   	volume = 284,
    	pages = {L9-L12},
      	doi = {10.1086/184341},
   	adsurl = {http://adsabs.harvard.edu/abs/1984ApJ...284L...9K}}

@ARTICLE{Planck_2016,
   author = {{Planck Collaboration} and {Ade}, P.~A.~R. and {Aghanim}, N. and 
	{Arnaud}, M. and {Ashdown}, M. and {Aumont}, J. and {Baccigalupi}, C. and 
	{Banday}, A.~J. and {Barreiro}, R.~B. and {Bartlett}, J.~G. and et al.},
    title = "{Planck 2015 results. XIII. Cosmological parameters}",
  journal = {\aap},
archivePrefix = "arXiv",
   eprint = {1502.01589},
     year = 2016,
    month = sep,
   volume = 594,
      eid = {A13},
    pages = {A13},
      doi = {10.1051/0004-6361/201525830},
   adsurl = {http://adsabs.harvard.edu/abs/2016A%26A...594A..13P}
}

@ARTICLE{Bryan_1998,
   author = {{Bryan}, G.~L. and {Norman}, M.~L.},
    title = "{Statistical Properties of X-Ray Clusters: Analytic and Numerical Comparisons}",
  journal = {\apj},
   eprint = {astro-ph/9710107},
     year = 1998,
    month = mar,
   volume = 495,
    pages = {80-99},
      doi = {10.1086/305262},
   adsurl = {http://adsabs.harvard.edu/abs/1998ApJ...495...80B}
}

@ARTICLE{Berlind_2002,
   author = {{Berlind}, A.~A. and {Weinberg}, D.~H.},
    title = "{The Halo Occupation Distribution: Toward an Empirical Determination of the Relation between Galaxies and Mass}",
  journal = {\apj},
   eprint = {astro-ph/0109001},
     year = 2002,
    month = aug,
   volume = 575,
    pages = {587-616},
      doi = {10.1086/341469},
   adsurl = {http://adsabs.harvard.edu/abs/2002ApJ...575..587B}
}

@ARTICLE{Sheth_Tormen_2004,
   	author = {{Sheth}, R.~K. and {Tormen}, G.},
    	title = "{On the environmental dependence of halo formation}",
  	journal = {\mnras},
   	eprint = {astro-ph/0402237},
     	year = 2004,
    	month = jun,
   	volume = 350,
    	pages = {1385-1390},
      	doi = {10.1111/j.1365-2966.2004.07733.x},
   	adsurl = {http://adsabs.harvard.edu/abs/2004MNRAS.350.1385S}
}

@ARTICLE{Gao_2005,
   	author = {{Gao}, L. and {Springel}, V. and {White}, S.~D.~M.},
   	 title = "{The age dependence of halo clustering}",
  	journal = {\mnras},
   	eprint = {astro-ph/0506510},
     	year = 2005,
    	month = oct,
   	volume = 363,
    	pages = {L66-L70},
      	doi = {10.1111/j.1745-3933.2005.00084.x},
   	adsurl = {http://adsabs.harvard.edu/abs/2005MNRAS.363L..66G}
}

@ARTICLE{Wechsler_2006,
   	author = {{Wechsler}, R.~H. and {Zentner}, A.~R. and {Bullock}, J.~S. and 
	{Kravtsov}, A.~V. and {Allgood}, B.},
    	title = "{The Dependence of Halo Clustering on Halo Formation History, Concentration, and Occupation}",
  	journal = {\apj},
   	eprint = {astro-ph/0512416},
     	year = 2006,
   	month = nov,
   	volume = 652,
    	pages = {71-84},
      	doi = {10.1086/507120},
   	adsurl = {http://adsabs.harvard.edu/abs/2006ApJ...652...71W}
}

@ARTICLE{Gao_White_2007,
   	author = {{Gao}, L. and {White}, S.~D.~M.},
    	title = "{Assembly bias in the clustering of dark matter haloes}",
  	journal = {\mnras},
   	eprint = {astro-ph/0611921},
     	year = 2007,
    	month = apr,
   	volume = 377,
    	pages = {L5-L9},
      	doi = {10.1111/j.1745-3933.2007.00292.x},
  	adsurl = {http://adsabs.harvard.edu/abs/2007MNRAS.377L...5G}
}

@ARTICLE{Salcedo_2018,
   author = {{Salcedo}, A.~N. and {Maller}, A.~H. and {Berlind}, A.~A. and 
	{Sinha}, M. and {McBride}, C.~K. and {Behroozi}, P.~S. and {Wechsler}, R.~H. and 
	{Weinberg}, D.~H.},
    title = "{Spatial clustering of dark matter haloes: secondary bias, neighbour bias, and the influence of massive neighbours on halo properties}",
  journal = {\mnras},
archivePrefix = "arXiv",
   eprint = {1708.08451},
     year = 2018,
    month = apr,
   volume = 475,
    pages = {4411-4423},
      doi = {10.1093/mnras/sty109},
   adsurl = {http://adsabs.harvard.edu/abs/2018MNRAS.475.4411S}
}

@ARTICLE{Zentner_et_al_2019,
       author = {{Zentner}, Andrew R. and {Hearin}, Andrew and {van den Bosch}, Frank C. and
         {Lange}, Johannes U. and {Villarreal}, Antonio},
        title = "{Constraints on assembly bias from galaxy clustering}",
      journal = {\mnras},
         year = "2019",
        month = "May",
       volume = {485},
       number = {1},
        pages = {1196-1209},
          doi = {10.1093/mnras/stz470},
archivePrefix = {arXiv},
       eprint = {1606.07817},
 primaryClass = {astro-ph.GA},
       adsurl = {https://ui.adsabs.harvard.edu/abs/2019MNRAS.485.1196Z}
}

@ARTICLE{Wang_Mo_Jing_2007,
   author = {{Wang}, H.~Y. and {Mo}, H.~J. and {Jing}, Y.~P.},
    title = "{Environmental dependence of cold dark matter halo formation}",
  journal = {\mnras},
   eprint = {astro-ph/0608690},
     year = 2007,
    month = feb,
   volume = 375,
    pages = {633-639},
      doi = {10.1111/j.1365-2966.2006.11316.x},
   adsurl = {http://adsabs.harvard.edu/abs/2007MNRAS.375..633W}
}

@ARTICLE{Li_Mo_Gao_2008,
   author = {{Li}, Y. and {Mo}, H.~J. and {Gao}, L.},
    title = "{On halo formation times and assembly bias}",
  journal = {\mnras},
archivePrefix = "arXiv",
   eprint = {0803.2250},
     year = 2008,
    month = sep,
   volume = 389,
    pages = {1419-1426},
      doi = {10.1111/j.1365-2966.2008.13667.x},
   adsurl = {http://adsabs.harvard.edu/abs/2008MNRAS.389.1419L}
}

@ARTICLE{Faltenbacher_White_2010,
   author = {{Faltenbacher}, A. and {White}, S.~D.~M.},
    title = "{Assembly Bias and the Dynamical Structure of Dark Matter Halos}",
  journal = {\apj},
archivePrefix = "arXiv",
   eprint = {0909.4302},
     year = 2010,
    month = jan,
   volume = 708,
    pages = {469-473},
      doi = {10.1088/0004-637X/708/1/469},
   adsurl = {http://adsabs.harvard.edu/abs/2010ApJ...708..469F}
}

@ARTICLE{Mao_Zentner_Wechsler_2018,
   author = {{Mao}, Y.-Y. and {Zentner}, A.~R. and {Wechsler}, R.~H.},
    title = "{Beyond assembly bias: exploring secondary halo biases for cluster-size haloes}",
  journal = {\mnras},
archivePrefix = "arXiv",
   eprint = {1705.03888},
     year = 2018,
    month = mar,
   volume = 474,
    pages = {5143-5157},
      doi = {10.1093/mnras/stx3111},
   adsurl = {http://adsabs.harvard.edu/abs/2018MNRAS.474.5143M}
}

@ARTICLE{Sato-Polito_et_al_2019,
       author = {{Sato-Polito}, Gabriela and {Montero-Dorta}, Antonio D. and
         {Abramo}, L. Raul and {Prada}, Francisco and {Klypin}, Anatoly},
        title = "{The dependence of halo bias on age, concentration, and spin}",
      journal = {\mnras},
         year = 2019,
        month = aug,
       volume = {487},
       number = {2},
        pages = {1570-1579},
          doi = {10.1093/mnras/stz1338},
archivePrefix = {arXiv},
       eprint = {1810.02375},
 primaryClass = {astro-ph.GA},
       adsurl = {https://ui.adsabs.harvard.edu/abs/2019MNRAS.487.1570S}
}

@ARTICLE{Xu_Zheng_2018,
   author = {{Xu}, X. and {Zheng}, Z.},
    title = "{Dependence of halo bias and kinematics on assembly variables}",
  journal = {\mnras},
archivePrefix = "arXiv",
   eprint = {1710.06862},
     year = 2018,
    month = sep,
   volume = 479,
    pages = {1579-1594},
      doi = {10.1093/mnras/sty1547},
   adsurl = {http://adsabs.harvard.edu/abs/2018MNRAS.479.1579X}
}

@ARTICLE{Zentner_Hearin_vdBosch_2014,
   author = {{Zentner}, A.~R. and {Hearin}, A.~P. and {van den Bosch}, F.~C.
	},
    title = "{Galaxy assembly bias: a significant source of systematic error in the galaxy-halo relationship}",
  journal = {\mnras},
archivePrefix = "arXiv",
   eprint = {1311.1818},
     year = 2014,
    month = oct,
   volume = 443,
    pages = {3044-3067},
      doi = {10.1093/mnras/stu1383},
   adsurl = {http://adsabs.harvard.edu/abs/2014MNRAS.443.3044Z}
}

@ARTICLE{Zheng_et_al_2005,
   author = {{Zheng}, Z. and {Berlind}, A.~A. and {Weinberg}, D.~H. and {Benson}, A.~J. and 
	{Baugh}, C.~M. and {Cole}, S. and {Dav{\'e}}, R. and {Frenk}, C.~S. and 
	{Katz}, N. and {Lacey}, C.~G.},
    title = "{Theoretical Models of the Halo Occupation Distribution: Separating Central and Satellite Galaxies}",
  journal = {\apj},
   eprint = {astro-ph/0408564},
     year = 2005,
    month = nov,
   volume = 633,
    pages = {791-809},
      doi = {10.1086/466510},
   adsurl = {http://adsabs.harvard.edu/abs/2005ApJ...633..791Z}
}

@Article{Hunter_2007,
  Author    = {Hunter, J. D.},
  Title     = {Matplotlib: A 2D graphics environment},
  Journal   = {Computing In Science \& Engineering},
  Volume    = {9},
  Number    = {3},
  Pages     = {90--95},
  publisher = {IEEE COMPUTER SOC},
  doi       = {10.1109/MCSE.2007.55},
  year      = 2007
}

@manual{GSL_2009,
  title        = {GNU Scientific Library Reference Manual}, 
  author       = {M. Galassi and J. Davies and J. Theiler and B. Gough and G. Jungman and P. Alken and M. Booth and F. Rossi},
  edition      = 3,
  month        = Jan,
  year         = 2009,
}

@misc{OhioSupercomputerCenter1987,
ark = {ark:/19495/f5s1ph73},
howpublished = {\url{http://osc.edu/ark:/19495/f5s1ph73}},
year  = {1987},
author = {{Ohio Supercomputer Center}},
title = {Ohio Supercomputer Center}
}

@ARTICLE{Zheng_Weinberg_2007,
   author = {{Zheng}, Z. and {Weinberg}, D.~H.},
    title = "{Breaking the Degeneracies between Cosmology and Galaxy Bias}",
  journal = {\apj},
   eprint = {astro-ph/0512071},
     year = 2007,
    month = apr,
   volume = 659,
    pages = {1-28},
      doi = {10.1086/512151},
   adsurl = {http://adsabs.harvard.edu/abs/2007ApJ...659....1Z}
}

@ARTICLE{Harker_et_al_2006,
   author = {{Harker}, G. and {Cole}, S. and {Helly}, J. and {Frenk}, C. and 
	{Jenkins}, A.},
    title = "{A marked correlation function analysis of halo formation times in the Millennium Simulation}",
  journal = {\mnras},
   eprint = {astro-ph/0510488},
     year = 2006,
    month = apr,
   volume = 367,
    pages = {1039-1049},
      doi = {10.1111/j.1365-2966.2006.10022.x},
   adsurl = {http://adsabs.harvard.edu/abs/2006MNRAS.367.1039H}
}

@ARTICLE{mmm15,
       author = {{More}, Surhud and {Miyatake}, Hironao and {Mandelbaum}, Rachel and {Takada}, Masahiro and {Spergel}, David N. and {Brownstein}, Joel R. and {Schneider}, Donald P.},
        title = "{The Weak Lensing Signal and the Clustering of BOSS Galaxies. II. Astrophysical and Cosmological Constraints}",
      journal = {\apj},
         year = 2015,
        month = jun,
       volume = {806},
       number = {1},
          eid = {2},
        pages = {2},
          doi = {10.1088/0004-637X/806/1/2},
archivePrefix = {arXiv},
       eprint = {1407.1856},
 primaryClass = {astro-ph.CO},
       adsurl = {https://ui.adsabs.harvard.edu/abs/2015ApJ...806....2M}
}

@ARTICLE{smm23,
       author = {{Sugiyama}, Sunao and {Miyatake}, Hironao and {More}, Surhud and {Li}, Xiangchong and {Shirasaki}, Masato and {Takada}, Masahiro and {Kobayashi}, Yosuke and {Takahashi}, Ryuichi and {Nishimichi}, Takahiro and {Nishizawa}, Atsushi J. and {Rau}, Markus M. and {Zhang}, Tianqing and {Dalal}, Roohi and {Mandelbaum}, Rachel and {Strauss}, Michael A. and {Hamana}, Takashi and {Oguri}, Masamune and {Osato}, Ken and {Kannawadi}, Arun and {Armstrong}, Robert and {Komiyama}, Yutaka and {Lupton}, Robert H. and {Lust}, Nate B. and {Miyazaki}, Satoshi and {Murayama}, Hitoshi and {Okura}, Yuki and {Price}, Paul A. and {Tait}, Philip J. and {Tanaka}, Masayuki and {Wang}, Shiang-Yu},
        title = "{Hyper Suprime-Cam Year 3 Results: Cosmology from Galaxy Clustering and Weak Lensing with HSC and SDSS using the Minimal Bias Model}",
      journal = {arXiv e-prints},
         year = 2023,
        month = apr,
          eid = {arXiv:2304.00705},
        pages = {arXiv:2304.00705},
          doi = {10.48550/arXiv.2304.00705},
archivePrefix = {arXiv},
       eprint = {2304.00705},
 primaryClass = {astro-ph.CO},
       adsurl = {https://ui.adsabs.harvard.edu/abs/2023arXiv230400705S}
}

@ARTICLE{mkt22,
       author = {{Miyatake}, Hironao and {Kobayashi}, Yosuke and {Takada}, Masahiro and {Nishimichi}, Takahiro and {Shirasaki}, Masato and {Sugiyama}, Sunao and {Takahashi}, Ryuichi and {Osato}, Ken and {More}, Surhud and {Park}, Youngsoo},
        title = "{Cosmological inference from an emulator based halo model. I. Validation tests with HSC and SDSS mock catalogs}",
      journal = {\prd},
         year = 2022,
        month = oct,
       volume = {106},
       number = {8},
          eid = {083519},
        pages = {083519},
          doi = {10.1103/PhysRevD.106.083519},
archivePrefix = {arXiv},
       eprint = {2101.00113},
 primaryClass = {astro-ph.CO},
       adsurl = {https://ui.adsabs.harvard.edu/abs/2022PhRvD.106h3519M}
}

@ARTICLE{mst23,
       author = {{Miyatake}, Hironao and {Sugiyama}, Sunao and {Takada}, Masahiro and {Nishimichi}, Takahiro and {Li}, Xiangchong and {Shirasaki}, Masato and {More}, Surhud and {Kobayashi}, Yosuke and {Nishizawa}, Atsushi J. and {Rau}, Markus M. and {Zhang}, Tianqing and {Takahashi}, Ryuichi and {Dalal}, Roohi and {Mandelbaum}, Rachel and {Strauss}, Michael A. and {Hamana}, Takashi and {Oguri}, Masamune and {Osato}, Ken and {Luo}, Wentao and {Kannawadi}, Arun and {Hsieh}, Bau-Ching and {Armstrong}, Robert and {Komiyama}, Yutaka and {Lupton}, Robert H. and {Lust}, Nate B. and {MacArthur}, Lauren A. and {Miyazaki}, Satoshi and {Murayama}, Hitoshi and {Okura}, Yuki and {Price}, Paul A. and {Sunayama}, Tomomi and {Tait}, Philip J. and {Tanaka}, Masayuki and {Wang}, Shiang-Yu},
        title = "{Hyper Suprime-Cam Year 3 Results: Cosmology from Galaxy Clustering and Weak Lensing with HSC and SDSS using the Emulator Based Halo Model}",
      journal = {arXiv e-prints},
         year = 2023,
        month = apr,
          eid = {arXiv:2304.00704},
        pages = {arXiv:2304.00704},
          doi = {10.48550/arXiv.2304.00704},
archivePrefix = {arXiv},
       eprint = {2304.00704},
 primaryClass = {astro-ph.CO},
       adsurl = {https://ui.adsabs.harvard.edu/abs/2023arXiv230400704M}
}

@ARTICLE{kfp21,
       author = {{Krause}, E. and {Fang}, X. and {Pandey}, S. and {Secco}, L.~F. and {Alves}, O. and {Huang}, H. and {Blazek}, J. and {Prat}, J. and {Zuntz}, J. and {Eifler}, T.~F. and {MacCrann}, N. and {DeRose}, J. and {Crocce}, M. and {Porredon}, A. and {Jain}, B. and {Troxel}, M.~A. and {Dodelson}, S. and {Huterer}, D. and {Liddle}, A.~R. and {Leonard}, C.~D. and {Amon}, A. and {Chen}, A. and {Elvin-Poole}, J. and {Fert{\'e}}, A. and {Muir}, J. and {Park}, Y. and {Samuroff}, S. and {Brandao-Souza}, A. and {Weaverdyck}, N. and {Zacharegkas}, G. and {Rosenfeld}, R. and {Campos}, A. and {Chintalapati}, P. and {Choi}, A. and {Di Valentino}, E. and {Doux}, C. and {Herner}, K. and {Lemos}, P. and {Mena-Fern{\'a}ndez}, J. and {Omori}, Y. and {Paterno}, M. and {Rodriguez-Monroy}, M. and {Rogozenski}, P. and {Rollins}, R.~P. and {Troja}, A. and {Tutusaus}, I. and {Wechsler}, R.~H. and {Abbott}, T.~M.~C. and {Aguena}, M. and {Allam}, S. and {Andrade-Oliveira}, F. and {Annis}, J. and {Bacon}, D. and {Baxter}, E. and {Bechtol}, K. and {Bernstein}, G.~M. and {Brooks}, D. and {Buckley-Geer}, E. and {Burke}, D.~L. and {Carnero Rosell}, A. and {Carrasco Kind}, M. and {Carretero}, J. and {Castander}, F.~J. and {Cawthon}, R. and {Chang}, C. and {Costanzi}, M. and {da Costa}, L.~N. and {Pereira}, M.~E.~S. and {De Vicente}, J. and {Desai}, S. and {Diehl}, H.~T. and {Doel}, P. and {Everett}, S. and {Evrard}, A.~E. and {Ferrero}, I. and {Flaugher}, B. and {Fosalba}, P. and {Frieman}, J. and {Garc{\'\i}a-Bellido}, J. and {Gaztanaga}, E. and {Gerdes}, D.~W. and {Giannantonio}, T. and {Gruen}, D. and {Gruendl}, R.~A. and {Gschwend}, J. and {Gutierrez}, G. and {Hartley}, W.~G. and {Hinton}, S.~R. and {Hollowood}, D.~L. and {Honscheid}, K. and {Hoyle}, B. and {Huff}, E.~M. and {James}, D.~J. and {Kuehn}, K. and {Kuropatkin}, N. and {Lahav}, O. and {Lima}, M. and {Maia}, M.~A.~G. and {Marshall}, J.~L. and {Martini}, P. and {Melchior}, P. and {Menanteau}, F. and {Miquel}, R. and {Mohr}, J.~J. and {Morgan}, R. and {Myles}, J. and {Palmese}, A. and {Paz-Chinch{\'o}n}, F. and {Petravick}, D. and {Pieres}, A. and {Plazas Malag{\'o}n}, A.~A. and {Sanchez}, E. and {Scarpine}, V. and {Schubnell}, M. and {Serrano}, S. and {Sevilla-Noarbe}, I. and {Smith}, M. and {Soares-Santos}, M. and {Suchyta}, E. and {Tarle}, G. and {Thomas}, D. and {To}, C. and {Varga}, T.~N. and {Weller}, J.},
        title = "{Dark Energy Survey Year 3 Results: Multi-Probe Modeling Strategy and Validation}",
      journal = {arXiv e-prints},
         year = 2021,
        month = may,
          eid = {arXiv:2105.13548},
        pages = {arXiv:2105.13548},
          doi = {10.48550/arXiv.2105.13548},
archivePrefix = {arXiv},
       eprint = {2105.13548},
 primaryClass = {astro-ph.CO},
       adsurl = {https://ui.adsabs.harvard.edu/abs/2021arXiv210513548K}
}

@ARTICLE{pkd22,
       author = {{Pandey}, S. and {Krause}, E. and {DeRose}, J. and {MacCrann}, N. and {Jain}, B. and {Crocce}, M. and {Blazek}, J. and {Choi}, A. and {Huang}, H. and {To}, C. and {Fang}, X. and {Elvin-Poole}, J. and {Prat}, J. and {Porredon}, A. and {Secco}, L.~F. and {Rodriguez-Monroy}, M. and {Weaverdyck}, N. and {Park}, Y. and {Raveri}, M. and {Rozo}, E. and {Rykoff}, E.~S. and {Bernstein}, G.~M. and {S{\'a}nchez}, C. and {Jarvis}, M. and {Troxel}, M.~A. and {Zacharegkas}, G. and {Chang}, C. and {Alarcon}, A. and {Alves}, O. and {Amon}, A. and {Andrade-Oliveira}, F. and {Baxter}, E. and {Bechtol}, K. and {Becker}, M.~R. and {Camacho}, H. and {Campos}, A. and {Carnero Rosell}, A. and {Carrasco Kind}, M. and {Cawthon}, R. and {Chen}, R. and {Chintalapati}, P. and {Davis}, C. and {Di Valentino}, E. and {Diehl}, H.~T. and {Dodelson}, S. and {Doux}, C. and {Drlica-Wagner}, A. and {Eckert}, K. and {Eifler}, T.~F. and {Elsner}, F. and {Everett}, S. and {Farahi}, A. and {Fert{\'e}}, A. and {Fosalba}, P. and {Friedrich}, O. and {Gatti}, M. and {Giannini}, G. and {Gruen}, D. and {Gruendl}, R.~A. and {Harrison}, I. and {Hartley}, W.~G. and {Huff}, E.~M. and {Huterer}, D. and {Kovacs}, A. and {Leget}, P.~F. and {McCullough}, J. and {Muir}, J. and {Myles}, J. and {Navarro-Alsina}, A. and {Omori}, Y. and {Rollins}, R.~P. and {Roodman}, A. and {Rosenfeld}, R. and {Sevilla-Noarbe}, I. and {Sheldon}, E. and {Shin}, T. and {Troja}, A. and {Tutusaus}, I. and {Varga}, T.~N. and {Wechsler}, R.~H. and {Yanny}, B. and {Yin}, B. and {Zhang}, Y. and {Zuntz}, J. and {Abbott}, T.~M.~C. and {Aguena}, M. and {Allam}, S. and {Annis}, J. and {Bacon}, D. and {Bertin}, E. and {Brooks}, D. and {Burke}, D.~L. and {Carretero}, J. and {Conselice}, C. and {Costanzi}, M. and {da Costa}, L.~N. and {Pereira}, M.~E.~S. and {De Vicente}, J. and {Dietrich}, J.~P. and {Doel}, P. and {Evrard}, A.~E. and {Ferrero}, I. and {Flaugher}, B. and {Frieman}, J. and {Garc{\'\i}a-Bellido}, J. and {Gaztanaga}, E. and {Gerdes}, D.~W. and {Giannantonio}, T. and {Gschwend}, J. and {Gutierrez}, G. and {Hinton}, S.~R. and {Hollowood}, D.~L. and {Honscheid}, K. and {James}, D.~J. and {Jeltema}, T. and {Kuehn}, K. and {Kuropatkin}, N. and {Lahav}, O. and {Lima}, M. and {Lin}, H. and {Maia}, M.~A.~G. and {Marshall}, J.~L. and {Melchior}, P. and {Menanteau}, F. and {Miller}, C.~J. and {Miquel}, R. and {Mohr}, J.~J. and {Morgan}, R. and {Palmese}, A. and {Paz-Chinch{\'o}n}, F. and {Petravick}, D. and {Pieres}, A. and {Plazas Malag{\'o}n}, A.~A. and {Sanchez}, E. and {Scarpine}, V. and {Serrano}, S. and {Smith}, M. and {Soares-Santos}, M. and {Suchyta}, E. and {Tarle}, G. and {Thomas}, D. and {Weller}, J. and {DES Collaboration}},
        title = "{Dark Energy Survey year 3 results: Constraints on cosmological parameters and galaxy-bias models from galaxy clustering and galaxy-galaxy lensing using the redMaGiC sample}",
      journal = {\prd},
         year = 2022,
        month = aug,
       volume = {106},
       number = {4},
          eid = {043520},
        pages = {043520},
          doi = {10.1103/PhysRevD.106.043520},
archivePrefix = {arXiv},
       eprint = {2105.13545},
 primaryClass = {astro-ph.CO},
       adsurl = {https://ui.adsabs.harvard.edu/abs/2022PhRvD.106d3520P}
}

@ARTICLE{Croton_et_al_2007,
       author = {{Croton}, Darren J. and {Gao}, Liang and {White}, Simon D.~M.},
        title = "{Halo assembly bias and its effects on galaxy clustering}",
      journal = {\mnras},
         year = 2007,
        month = feb,
       volume = {374},
       number = {4},
        pages = {1303-1309},
          doi = {10.1111/j.1365-2966.2006.11230.x},
archivePrefix = {arXiv},
       eprint = {astro-ph/0605636},
 primaryClass = {astro-ph},
       adsurl = {https://ui.adsabs.harvard.edu/abs/2007MNRAS.374.1303C}
}
\bibliographystyle{aasjournal}

\label{lastpage}

\end{document}